\documentclass[prd,reprint]{revtex4}                                                   
\usepackage[pdftex]{graphicx}
\usepackage{wrapfig, epsfig}

\newcommand{\keyword}[1]{\textsf{\slshape #1}}

\begin{document}
\title{Feebly Self-Interacting Cold Dark Matter:\\
New theory for the Core-Halo structure in GLSB Galaxies}
\author{Himanshu Kumar}
\email{hman_19@hotmail. com}
\author{Sharf Alam}
\affiliation{Jamia Millia Islamia,  New Delhi,  India}

\begin{abstract}
We explore the low energy cosmological dynamics of feebly self-interacting cold dark matter and propose a new simple explanation for the rotation curves of the core-halo model in massive LSB (Low Surface brightness)galaxies. We argue in favor of the truly collisionless nature of cold
dark matter,which is feebly,self-interacting at small scales between epochs of equality and recombination.For this,we assume
a model, wherein strongly coupled baryon-radiation plasma ejects out
of small regions of concentrated cold dark matter without losing its
equilibrium.  We use the Merscerskii equation i.e. the variable mass formalism of classical dynamics.We obtain new results relating the oscillations in the CMB anisotropy to the ejection velocity of the baryon-radiation plasma,which can be useful tool for numerical work for exploring the second peak of CMB.
Based on this model, we discuss the growth of perturbations in such a feebly self-interacting,cold dark
matter both in the Jeans theory and in the expanding universe using
Newton's theory.We obtain an expression for the growth of fractional perturbations in cold dark matter,which reduce to the standard result of perturbation theory for late recombination epochs.  We see the effect of the average of the perturbations in the cold
dark matter potential on the cosmic microwave background temperature anisotropy that
originated at redshifts between equality and recombination i.e. $1100 < z < z_{eq}$.  Also we obtain an expression for the Sachs-Wolfe effect,i.e. the CMB temperature anisotropy at decoupling in terms of the average of the perturbations in
cold dark matter potential.
\end{abstract}
\maketitle

\keyword{Self-interacting,Cold Dark
matter,Core-Cusp,Dark matter halos, Collisionless,WIMP,Rotation Curves,LSB Galaxies, Baryon-Radiation plasma,Dark Energy,Equality,Recombination,Decoupling, ,Perturbations,Jeans Theory, CMB, Sachs-Wolfe effect,Merscerskii equation}

\section{Introduction}
\label{sec:1 Intro}
Flat cosmological models with a mixture of ordinary baryonic matter, cold
dark matter, and cosmological constant(or quintessence) and a nearly
scale-invariant, adiabatic spectrum of density fluctuations are consistent
with standard inflationary cosmology.  They provide an excellent fit to
current observations on large scales($>>1$ Mpc).  Currently, the constitution
of the universe is 4\% baryons, 23\% dark matter and 73\% dark energy
\cite{Reid2010,Percival2010,Huff,Tinker2012,Larson2011,Dunkey2011}. 

In the standard hot Big Bang model, the universe
is initially hot and the energy density is dominated by radiation.  The
transition to matter domination occurs at $z\approx 10^{4}$.  In the
epochs after equality and before recombination, the universe remains
hot enough.  Thus the gas is ionized, and the electron-photon scattering
effectively couples the matter and radiation \cite{Peebles1970}.  At
$z\approx 1200$, the temperature drops below $\approx 3300$ K.  The
protons and electrons now recombine to form neutral hydrogen
and neutral helium.  This event is usually known as recombination
\cite{Peebles1968,Zeldovich1969,Seager2000}.  The
photons then decouple and travel freely.  These photons which
keep on travelling till present times are observed as the cosmic
microwave background(CMB).  The cold dark matter theory including
cosmic inflation is the basis of standard modern cosmology. \\
\indent  This model ,which we shall call "Standard"$\Lambda$CDM,or s$\Lambda$CDM for short is consistent with many fundamental cosmological observations:the age of the universe as measured from the oldest stars \cite{Chaboyer1998}),the extragalactic distance scale as
measured by distant Cepheids  \cite{Madore1998}); the primordial abundance of the light
elements \cite{Schramm1998}), the baryonic mass fraction of galaxy clusters \cite{White1993}, the amplitude of the Cosmic Microwave Background fluctuations measured by COBE
(\cite{Lawrence1999}\cite{PJE1982}, the present-day abundance of massive galaxy clusters  \cite{Eke1996} ), the shape and amplitude of galaxy clustering patterns \cite{Rees1999}), the magnitude of large-scale coherent motions of galaxy systems  \cite{Strauss1995}, \cite{Zaroubi1997} ), and the world geometry inferred from observations of distant type Ia supernovae \cite{Perl1999} ,  \cite{Garnavich1998}),large scale structure data \cite{Pagels1982} among others.Moreover,this model assumes the dominance of non-baryonic matter \cite{Tegmark2004,Spergel2007},which is massive,weakly interactive and collisionless \cite{BondSzalay1983} and has negligibly small primeval velocity dispersion together with electromagnetic neutrality \cite{Sigurdson2004,Mcdermott2011}.The word Cold means that the thermal motions of DM
particles were essentially negligible at the time of matter-radiation
equality i.e.
the ratio $\frac{T}{M}<\phi$,where $\phi$ is the gravitation potential and T and M
represent the temperature and mass of the dark matter particle respectively.Though,s$\Lambda$CDM
predicts and matches many available data fairly well \cite{Efstathiou1990,Ostriker2002} and predicts a lot about structure formation \cite{Peebles1980,White2010},but at small scales,the core-cusp issue is one of the major problems which cannot be explained by s$\Lambda$CDM.The origin of this problem lies in the fact that there is no mechanism in s$\Lambda$CDM which can explain a very steeply rising rotation curves from CDM  simulations as compared to slowly rising ones from observational data 
\cite{Pickering1997},\cite{W.J.G.1996}.There is a need to explain the positive slopes in the outer parts of rotation curves of LSB galaxies,upto the outermost measured point,especially more positive than the HSB galaxies \cite{W.J.G.1996}.Either the picture suggested by CDM simulations is wrong or we need new physics under the CDM paradigm to resolve the core-cusp issue i.e. to try to explain the observed rotation curves of DM dominated galaxies (LSBs)\cite{Barker2003}\cite{KuziodeNaray2007}.   \\
The existence of clusters($\leq 50$ mpc) and groups of galaxies suggests
that the galaxy formation is due to the gravitational instability
of a spatially homogenous and isotropic expanding universe.  The
Perturbations of such a model have been first investigated by
\cite{Jeans1902}.  Then within the framework
of general relativity by \cite{Lifshitz1946,Khalatnikov1963}.  Also
it was studied in a newtonian model\cite{Bonnor1957}.  The results
obtained from relativistic theory were similar to the standard
Jeans theory.  In the early universe, radiation fixes the expansion
rate.  This is due to low density and self-gravity of dark matter
\cite{Guyot1970,Meszaros1974}.\\
  \indent
  We know that the dark matter reveals itself only through its gravitational interactions.Except some recent astrophysical observations,there are no observational facts that indicate existence of significant interactions between the ordinary and dark matter particles.The observations of the Bullet cluster,more formally known as 1E0656-56 \cite{Clowe2006} and other merging clusters CL 0152-1357 \cite{Jee2005} and MS 1054-0321 \cite{Ford2005} prove that the dark matter components of the colliding clusters do not perturb each other much during the collision, allowing themselves to pass right through.Independent results on the dark matter cross-section supporting the collisionless nature have been reported from the studies of A2744 \cite{Merten2011},MACS J0025. 4-1222 \cite{Bradac2008}.There are several other merging clusters that provide results similar to the Bullet cluster A1758 \cite{Okabe2008},\cite{Ragozzine2012},A2163 \cite{Soucail2012}.That is these observations support the WIMP based dark matter models \\
  \indent
We know that the self-interacting nature of cold dark matter has been supported by some recent observational data.The merging cluster of galaxies have the power to directly probe and place limilts on the self-interaction cross-section of dark matter \cite{Markevitch2004} .This has been used to place upper limits on the dark matter particle self-interaction cross-section of the order of 1 $cm^{2}gm^{-1}$.If the dark matter has a large self-interaction cross-section,then it would not be located in the vicinity of the cluster galaxies.Williams and Saha have suggested that a kpc scale separation between stellar and dark matter components in the cluster A3827 may be evidence for the dark matter with a non-negligible self-interaction cross-section \cite{Williams2011}.A large self-interaction cross-section for darkmatter much beyond the upper limit on the cross-section derived from the Bullet cluster has been derived by Randal et al \cite{Randall2008}.A higher limit on the DM self-interacting cross-section $\sigma_{DM}m^{-1}\lesssim 7 cm^{2}gm^{-1}$ has been derived in \cite{Dawson2011}.The galaxies outrunning the dark matter in the Musket Ball cluster DLSCL J0916.2+2951 suggests that dark matter particles do interact and slow down like a gas.The self-ineracting nature of dark matter has been proposed as a possible solution for the shallow core of the halos \cite{Steinhardt2000}.This excludes a very strong self-interacting DM, as this could lead to gravothermal catastrophe with a very steep central density profile in galaxies \cite{Firmani2000}.\\
\indent In this paper,we strongly argue in favor of the collisionless nature of cold  dark matter particles,which are feebly self-interacting at very small scales.This we do by an assumption inherent in the model, which ensures that the dark matter is collisionless for all epochs upto recombination.This is an improvement over the standard perturbation theory in the static and the Newtonian expanding scenario.We incorporate the feebly,self-interacting nature by allowing the cold dark matter to feel the pressure of the baryon-radiation fluid,at very small scales.This is because of the very heavy nature of cold dark matter particles and due to minor fluctuations in it's velocity as a result of self-interactions. \\
\indent
 After equality, the main contribution
to the gravitational potential is due to the cold dark matter.Therefore in the metric of the perturbed FRW universe, the main contribution to
the potential comes from an imperfect fluid, i. e.  the dark
matter.  We consider that the difference $\psi - \phi $ is very small as
compared to $ \psi $.The ratio of   $\psi -\phi$ to $\psi$ is proportional to the ratio of the photon mean free path to
the perturbation scale.  Here $\psi$ and $\phi$ are the newtonian potential
and the perturbations to the spatial curvature in a conformal Newtonian
gauge respectively(see eq. 48 sec~\ref{sec:2}).  We neglect the contribution
of the baryon density in realistic models,  where baryon contributes only
a small fraction of the total matter density.The potential of cold dark matter is time-independent both
for long wavelength and short wavelength perturbations.This is because it is highly non-relativistic.  There is a strong coupling between the
baryon and the radiation in the plasma.So, we treat it as a single perfect fluid for low baryon densities.  We neglect the non-diagonal components in the energy- momentum tensor of dark matter.  This is because for very small photon mean free paths  $\psi$ is same as $\phi$.  Therefore, we treat it as a single perfect fluid for many epochs after equality
upto the recombination.  It is only after recombination,  that the baryon
density starts increasing.  This is after the primordial nucleosynthesis
of hydrogen and helium is complete.  The radiation after decoupling from
matter at very late recombination epochs evolves separately.  The CMB
anisotropy of epochs between equality and recombination gives important
information about the perturbations in the cold dark matter i. e.   the
modes that enter the horizon before recombination.\\
\indent
 We assume a model,based on the variable mass formalism,inspired from the Merscerskii equation of classical dynamics.Here,the strongly coupled baryon-radiation plasma ejects out
of a concentrated region of cold dark matter.In the classical Merscerskii problem it is the relative velocity of escaping 
(or incident)mass with respect to the center of mass of the body (which is at rest...).We can use this  model here for a small volume element $\Delta V$ filled in by the baryon-photon fluid,whose skeletal structure is composed of dark matter .This volume element is of the order of $\sim 50$ kpc.The outermost skeletal region extending for $\approx 5$ kpc has exponentially decreasing density profile.Due to very heavy and ultra non-relativistic nature of CDM, the center of mass of the volume element will be at rest for an anisotropic ejection of baryon-photon plasma.In the standard Einstein static case of jeans theory or in the expanding Newtonian one,we cannot appreciate the feeble self-interactions triggered in the cold dark matter particles,at very small scales,due to the flow of strongly coupled baryon-photon plasma.The investigation of the perturbation theory using the variable mass formalism gives an opportunity to see the relative transfer of energy and momentum between cold dark matter particles,especially if they can self-interact.Also,in the standard Newtonian approach or in the Jeans theory,we do not have a method to preserve the collisionless nature of cold dark matter,throughout the evolution of perturbations in the matter,with its various substituents.Here,we can preserve  the collisionless nature of cold dark matter particles throughout and still recover the standard CMB.Moreover,we have a new physics to explain the core-halo structure of massive,DM dominated (LSB) galaxies.\\ 
\indent  
  The picture that the  dark matter forms the skeletal structure,which is filled in by other matter is supported from latest cosmological observations\cite{Massey2007}
  \cite{feng2010}.Also,we know that the baryonic matter collapses in the gravitational potential of dark matter.This implies that that there will be regions of dark matter across which there will be a flow of baryonic matter and radiation.Therefore,we need a model,which can emphasize the flow of strongly coupled baryon-radiation plasma through any volume element formed by the skeletal structure of dark matter.This could help us in understanding the growth of velocities and internal masses of galaxies with distance from the galactic centre and thus try to resolve the core-cusp issue in LSB galaxies.\\
  \indent We know that there are only two major components of universe between epochs of equality and recombination,i.e. the strongly coupled baryon-radiation fluid under the influence of the gravitational potential of cold dark matter.Since,we assume a picture of a volume element with its skeletal structure formed from cold dark matter and the strongly coupled baryon-radiation fluid flowing through it,we are inspired to study this two-component system from variable mass dynamics.This is a formalism of classical dynamics which is used to study the evolution of the skeletal structure of the rocket,because of the gases ejecting out of it. Based on this model,we
study the dynamics of the cold dark matter after equality and upto many
epochs near recombination.  Note that this sort of flow assumes that the
baryon -radiation plasma ejects out without losing its equilibrium in
the presence of cold dark matter.  This is possible only if we assume
that the cold dark matter is truly collisionless,but it can be feebly self-interacting.This we do by allowing the cold dark matter to feel feeble pressure,triggered by its own self-interactions,and also because of its very heavy nature,at very small scales.This we do by ensuring that the baryon -radiation plasma obeys an equilibrium equation for all epochs between equality and recombination.  This assumption can also
accomodate a feebly,self-interacting dark matter at very small scales, which can
transfer energy and momentum to the outer core \cite{Steinhardt2000}. \\
\indent
We discuss the dynamics of feebly,self-interacting DM in the variable mass formalism with a purpose to explore a new physical explanation under the CDM paradigm for the rotation curves of massive, DM dominated galaxies (LSBs).Also,we explore this formalism,to bring in more parameters sensitive to the baryon-to-photon ratio in the CMB , e.g. $u_{0}$: the ejection velocity of the baryon-radiation fluid with respect to the concentrated region of cold dark matter and $c^{2}_{s}$:the square of the velocity of sound.This we know can be used to get more information about the second and higher peaks of CMB,through numerical analysis.The
strongly-coupled baryon -radiation plasma is only under the influence
of the potential of cold dark matter at these epochs.  We assume that the
entropy of photons relative to Cold dark matter is initially spatially uniform
on supercurvature scales(wavelengths greater than $ H^{-1}$).  During
this time, the matter and radiation densities vary in space.  In other
words, we consider adiabatic perturbations.  As the universe expands, the
inhomogeneity scale becomes smaller than the curvature scale.  Thus the
components move with respect to one another and the entropy of photon
relative to cold dark matter particle starts varying spatially.  In contrast, the entropy of photons relative to
 baryons remains spatially uniform on all scales.  This will continue until the
baryons decouple from radiation.\\
\indent
 We use Merscerskii equation to study the dynamics of the strongly
coupled baryon- radiation plasma, after the cold dark matter starts
to dominate.  This occurs after epochs of equality.Here we
imagine a flow of the baryon - radiation plasma across regions of highly
concentrated non -relativistic cold dark matter.We accomodate the self-interactions of ultra non-relativistic cold dark matter by letting it feel some pressure due to its heavy nature and minor fluctuations,though it does not itself contribute to pressure.The term $\Delta M_{d}\frac{d\vec{v_{d}}}{dt}$ in (Eq. 2)
sec.~\ref{sec:2} represents the minor fluctuations in the velocity of cold dark matter particles due to self-interactions.  The baryon -radiation
plasma can be assumed to have an ejection velocity with respect to the
cold dark matter after equality.  We do not disturb the equilibrium of
the ejecting baryon-radiation plasma, even in the presence of cold dark
matter.  This we do to argue in favour of truly collisionless nature of
cold dark matter.  We ensure this by a specific assumption(see eq. 27
sec. ~\ref{sec:2})which is valid for all epochs between equality and recombination,as an improvement over the standard perturbation theory.  This is the time,  when the radiation densities
rapidly start to decrease.  This is due to the fact that its scaling is proportional to
$a^{-4}$.The scaling of the cold dark matter density is proportional to $a^{-3}$.  On the basis
of this model,  we  study the adiabatic perturbations between epochs of
equality and recombination.  This we do, both in the Jean's theory and in
an expanding universe in Newtonian theory.  We then find out the effect
of perturbations in cold dark matter potential on the anisotropy in
temperature of radiation at these epochs.  This we do for modes greater
than the curvature scales.  These are the modes which enter horizon
before recombination.Finally,we use certain assumptions inherent in the model (see eq.36 sec.~\ref{sec:2} and eq.110 sec.~\ref{sec:3} to resolve the core-cusp problem in small-scale s$\Lambda$CDM cosmology using DM dominated galaxies as ideal test cases.In particular,we try to explain the rotation curves of LSB galaxies. \\
\indent
The paper proceeds as follows.  In section II, we discuss the Dynamics
of Cold dark matter in baryon-radiation plasma using the Merscerskii
equation in the Jeans theory.We ensure the collisionless nature of cold dark matter throughout, by an assumption inherent in the model.Then we discuss the adiabatic perturbations
in the scope of this model.We suggest the growth of dark energy with negative pressure at recombination epochs,which cancels the force due to pressure of baryon-radiation fluid (see eq.(32) 
sec. ~\ref{sec:2}). We also deduce an equation which represents
how the anisotropy in the temperature of radiation is affected by the
perturbations in the cold dark matter in the epochs between equality
and recombination.  We derive an expression for the Sachs-Wolfe effect.  In
Section III, in the scope of this model, we discuss the dynamics of the above
mentioned scenario in the expanding universe in the Newtonian theory.We keep the cold dark matter collisionless throughout by an inherent assumption.We
show how the anisotropy in the temperature of radiation is affected
by the perturbations in Cold dark matter potential,  in an expanding
universe.This depends on $u_{o}$,i.e. the ejection velocity of the baryon-photon fluid,a parameter sensitive to baryon-to-photon ratio,in addition to $c^{2}_{s}$,in the traditional approach   (see eq.(128),eq. (119)$\&$ eq.(133)  
sec. ~\ref{sec:2}) .  Also we evaluate an expression for the Sachs-Wolfe effect in
the expanding universe.  We then write the equation for the evolution of
cold dark matter perturbations with time.Finally,in sec.~\ref{sec:4},based on one of our assumptions (eq.36 sec.~\ref{sec:2} and eq. 110 sec.~\ref{sec:3})which are important in deriving the CMB in our model,we propose a new physical explanation under the paradigm of s$\Lambda$CDM to resolve the core-cusp controversy for galaxies in general,with regard to massive,LSB(DM dominated)galaxies.In particular,we try to explain the outer rotation curves of the massive,LSB galaxies and the linearly increasing mass of galaxies upto the last measured point \cite{Rubin1978}.

\section{Gravitational Instability: Jeans theory Representation}
\label{sec:2}
In Jeans theory we consider a static,  non-expanding universe.  Also we
assume a homogeneous, isotropic background with constant time -independent
matter density \cite{Jeans1902}.  This
assumption is in  obvious contradiction to the hydrodynamical equations.  In
fact, the energy density remains unchanged only if the matter is at
rest and the gravitational force, $F \propto \bigtriangledown\phi $
vanishes.  This inconsistency can in principle be avoided if we consider
a static Einstein universe, where the gravitational force  of the matter
is compensated by the antigravitational force of an appropriately chosen
cosmological constant. 
We consider the fundamental equation of dynamics of a mass point with variable mass.  This is also referred to as the Merscerskii equation. 
\begin{equation}
 m\frac{\vec{dv}}{dt} = \vec{F} + \frac{dm}{dt}\vec{u} 
\end{equation}
It should be pointed out that in an inertial frame,  $\vec{F}$ is
interpreted as the force of interaction of a given body with surrounding
bodies.  The last term          $\frac{dm}{dt}\vec{u}$ is referred to as
the reactive force $\vec{R}$.  This force appears as a  result of the action that
the added or separated mass exerts on a given body.  If mass is added
, then the $\frac{dm}{dt}>0$ coincides with the vector $\vec{u}$.  If mass
is separated, $\frac{dm}{dt}< 0$,the vector $\vec{R}$ is oppositely
directed to the vector $\vec{u}$. 

 We consider a fixed volume element $\Delta V$ in Euler (non-co-moving)
 co-ordinates $\vec{x}$.  After equality, when the cold dark matter starts
 to dominate in small regions of space, we write:
\begin{equation}
\Delta\, M_{d}\frac{d\vec{v_{d}}}{dt}= -\Delta M_{d}\cdot \vec{\nabla} \phi - \vec{\nabla} p_{b\gamma}\cdot \Delta V
- \frac{dM_{b\gamma}}{dt}\vec{u}
\end{equation}
Here the first term on L. H. S represents the acceleration in the mass of
cold dark matter of mass $\Delta M_{d}$ and $\vec{v_{d}}$ is the velocity
of dark matter element.We can write this term as we allow the dark matter to feel the pressure of the baryon-radiation fluid at small scales.This we assume due to very heavy nature of dark matter and due to minor fluctuations in its velocity because of self-interactions. The first term on the R. H. S.  represents the
gravitational force on the cold dark matter.  The second term on right
is for the force due to the pressure of the baryon -radiation plasma
\cite{Mukhanov2005}.  The last term on R. H. S.  is the reactive force
on the cold dark matter due to the ejection of baryon - radiation plasma
from regions dominated by cold dark matter.This ejection is not isotropic.This ejection has time and spatial dependence,so that the reactive force does not vanish. This last force is only due to the model which we assume here.  Here $p_{b\gamma}$ is the pressure
of the baryon-radiation plasma,and $\vec{u}$ is the ejection velocity of
the Baryon - radiation plasma with respect to the concentrated region of
cold dark matter.  Note that in (eq.2), we do not take the pressure of
cold dark matter into account.  This is because the cold dark matter
is highly non - relativistic.  Therefore,  we neglect its pressure i.e.($\vec{\nabla} p_{d}$ term). 
We assume that after equality, the strongly coupled Baryon - radiation
plasma starts to decouple from the matter.  The matter at these
epochs is predominantly Cold Dark matter.\\
\indent
After Equality, in regions dominated by highly non - relativistic Cold dark matter, we write the continuity equation for the ejection of baryon- radiation plasma complex :
\begin{equation}\
\frac{dM_{b\gamma}}{dt}=\int _{\Delta V} \frac{\partial \varepsilon_{b\gamma}}{dt}dV
\end{equation}
where $M_{b\gamma}$ is the ejecting mass of baryon-radiation plasma and
$ \varepsilon_{b\gamma}$
is the energy density of baryon radiation plasma in the concentrated
region of cold dark matter, from where it is ejecting out.  In this model, 
we assume that     the ejection of baryon -radiation plasma out of
a concentrated region of heavy Cold dark matter does not disturb the
equilibrium of the dark matter.  This  we can assume  only because of the
truly collisionless nature of Cold dark matter.  This is the fundamental
assumption of this model.  We neglect the flux of the Cold dark matter out
of a region of volume $\Delta V$ in the time that the baryon -radiation
plasma flows out of this region.The cold dark matter acts as a skeletal structure for the small volume element $\Delta V$.This explains the absence of time derivative of $\Delta V$ in eq. (4).The rate of flow is entirely determined
by the flux of the baryon - radiation plasma and   we write :
\begin{equation}
\frac{dM_{b\gamma}}{dt}= \int _{\Delta V} \frac{\partial \varepsilon_{b\gamma}}{dt}dV = - \int _{\Delta V}\vec{\nabla}\cdot \varepsilon _{b\gamma}\vec{u}dV
\end{equation}
 We  write (Eq. 2) as:
\begin{equation}
\varepsilon_{d}\Delta V\frac{d\vec{v_{d}}}{dt}= -\varepsilon_{d}\Delta V\vec{\nabla}\phi - \vec{\nabla} p_{b\gamma}\Delta V - \frac{\partial\varepsilon_{b\gamma}}{\partial t}\Delta V\vec{u}
\end{equation}
where $\varepsilon_{d}$ is the energy density of cold dark matter, and $
\phi$ is the gravitational potential of cold dark matter. 
\begin{equation}
\varepsilon_{d}[\frac{\partial\vec{v_{d}}}{\partial t} + \vec{v_{d}}\cdot \nabla\vec{v_{d}}]= -\varepsilon_{d}\vec{\nabla}\phi -\vec{\nabla} p_{b\gamma} - \frac{\partial\varepsilon_{b\gamma}}{\partial t}\vec{u}
\end{equation}
Now using the Jeans theory, we introduce small perturbations about the
equilibrium values of variables, W. Bonnor\cite{Bonnor1957}:
\begin{equation}
\varepsilon_{d}(\vec{x}, t)=\varepsilon_{do} + \delta\varepsilon_{d}(\vec{x}, t)
\end{equation}
\begin{equation}
\vec{v_{d}}(\vec{x}, t)= \vec{v}_{do} + \delta\vec{v_{d}}(\vec{x}, t)= \delta\vec{v}_{d}(\vec{x}, t)
\end{equation}
\begin{equation}
\phi(\vec{x}, t)= \phi_{o}+\delta\phi (\vec{x}, t)
\end{equation}
\begin{equation}
\vec{u}(\vec{x}, t)= \vec{u_{o}} + \delta\vec{u}(\vec{x}, t)
\end{equation}
where $\vec{v_{do}}<<c$, the speed of light, $\vec{u_{o}}\neq 0$ and $\delta\varepsilon_{d}<< \varepsilon_{do}$ 
\begin{equation}
p_{b\gamma}(\vec{x}, t)= p_{b\gamma}(\varepsilon_{b\gamma o} + \delta\varepsilon_{b\gamma}, S_{o} + \delta S) = p_{b\gamma o} + \delta p_{b\gamma}(\vec{x}, t)
\end{equation}
\begin{equation}
S(\vec{x}, t)= S_{o} + \delta S(\vec{x}, t)
\end{equation}
where S is the entropy of cold dark matter element.  Also we write:
\begin{equation}
\delta p_{b\gamma} = c_{s}^{2}\delta\varepsilon_{b\gamma} + \sigma\delta S
\end{equation}
where $c^{2}_{s}$ is the speed of sound.  Neglecting dissipation, we write:
\begin{equation}
\frac{dS}{dt} = \frac{\partial S}{\partial t} +  \vec{v_{d}}\cdot \vec{\nabla}S
\end{equation}
\begin{equation}
\nabla^{2}\phi = 4\pi G\varepsilon_{d}
\end{equation}
From (eq. 4) we write:
\begin{equation}
\frac{\partial\varepsilon_{b\gamma}}{\partial t} + \vec{\nabla}\cdot \varepsilon_{b\gamma}\vec{u} = 0
\end{equation}
We substitute the values from (eq. 7)-(eq. 15)in (eq. 16)and (eq. 6) to get:
\begin{equation}
\frac{\partial\delta\varepsilon_{b\gamma}}{\partial t} + \varepsilon_{b\gamma o}\vec{\nabla}\cdot \delta\vec{u} + \vec{\nabla}\cdot (\delta\varepsilon_{b\gamma}\vec{u_{o}}) = 0
\end{equation}
\begin{equation}
\varepsilon_{do}\frac{\partial\delta\vec{v_{d}}}{\partial t} = - \delta\varepsilon_{d}\nabla\phi_{o} - \nabla(c_{s}^{2}\delta\varepsilon_{b\gamma} + \sigma\delta S_{b\gamma}) - \frac{\partial\varepsilon_{b\gamma 0}}{\partial t}\delta\vec{u}  - \frac{\partial\delta\varepsilon_{b\gamma}}{\partial t}\vec{u_{o}}
\end{equation}
\begin{equation}
\nabla^{2}\delta\phi = 4\pi G\delta\varepsilon_{d}
\end{equation}
\begin{equation}
\frac{d\delta S}{dt} = \frac{\partial\delta S}{dt} + (\delta\vec{v_{d}}\cdot \nabla)S_{o} = 0 
\end{equation}
We now take the divergence of (eq. 18)to get:
\begin{eqnarray}
\varepsilon_{do}\frac{\partial \vec{\nabla}\cdot \delta\vec{v_{d}}}{\partial t} = - (\vec{\nabla}\cdot \delta\varepsilon_{d})\nabla\phi_{o} - \delta\varepsilon_{d}(4\pi G\varepsilon_{do}) 
-c_{s}^{2}\nabla^{2}\delta\varepsilon_{b\gamma}\nonumber \\
 - \sigma\nabla^{2}\delta S  - \vec{\nabla}\cdot \lbrace\frac{\partial(\varepsilon_{b\gamma o}\delta\vec{u})}{\partial t} - \varepsilon_{b\gamma o}\frac{\partial\delta\vec{u}}{\partial t}\rbrace - \vec{\nabla}\cdot \lbrace \frac{\partial (\delta\varepsilon_{b\gamma }\vec{u_{o}})}{\partial t} - \delta\varepsilon_{b\gamma}\frac{\partial \vec{u_{o}}}{\partial t}\rbrace  
\end{eqnarray}
Now using (eq. 17) and (eq. 19) we get :
\begin{equation}\frac{\partial^{2}\delta\varepsilon_{b\gamma}}{\partial t^{2}} - c_{s}^{2}\nabla^{2}\delta\varepsilon_{b\gamma} - 4\pi G \varepsilon_{do}\delta\varepsilon_{d} - (\vec{\nabla}\cdot \delta\varepsilon_{d})\nabla\phi_{o}
 -\varepsilon_{do}\frac{\partial(\vec{\nabla}\cdot \delta\vec{v_{d}})}{\partial t}  + \vec{\nabla}\cdot \lbrace\varepsilon_{b\gamma o}\frac{\partial\delta\vec{u} }{\partial t} + \delta\varepsilon_{b\gamma}\frac{\partial\vec{u_{o}}}{\partial t}\rbrace = \sigma\nabla^{2}\delta S
\end{equation}
In a static universe the total energy remains constant.  So for small inhomogeneities we write:
\begin{equation} \delta\varepsilon_{b\gamma} = -\delta\varepsilon_{d}\end{equation}
Thus (eq. 22) becomes:
\begin{equation}\frac{\partial^{2}\delta\varepsilon_{b\gamma}}{\partial t^{2}} - c_{s}^{2}\nabla^{2}\delta\varepsilon_{b\gamma} + 4\pi G\varepsilon_{do}\delta\varepsilon_{b\gamma} +(\vec{\nabla}\cdot \delta\varepsilon_{b\gamma})\nabla\phi_{o} 
-\varepsilon_{do}\frac{\partial\vec{\nabla}\cdot \vec{v_{d}}}{\partial t} + \vec{\nabla}\cdot \lbrace\varepsilon_{b\gamma o}\frac{\partial\delta\vec{u}}{\partial t} + \delta\varepsilon_{b\gamma}\frac{\partial\vec{u}}{\partial t}\rbrace
= \sigma\nabla^{2}S(x)
\end{equation}
For strongly -coupled baryon - radiation plasma before recombination, with low baryon densities we  write :
\begin{equation}\varepsilon_{b\gamma}+ p_{b\gamma}=\varepsilon_{b}
+ \frac{4}{3}\varepsilon_{\gamma} = \frac{4}{9c_{s}^{2}}\varepsilon_{\gamma}\end{equation}
Thus we get:
\begin{equation}
\delta\varepsilon_{b\gamma} + \delta p_{b\gamma} = \frac{4}{9c_{s}^{2}}\delta\varepsilon_{\gamma}
\end{equation}
The baryon - radiation plasma is only affected by the gravitational
potential of the cold dark matter in these epochs.  We argue that the
baryon-radiation plasma ejecting out of concentrated regions of cold
dark matter is always in equilibrium, even in the presence of cold
dark matter.  This is possible only if the cold dark matter is truly
collisionless.  So for epochs between equality and recombination, we write:
\begin{equation}
 \frac{\partial p_{b\gamma}}{\partial T} = \frac{\varepsilon_{b\gamma} + p_{b\gamma}}{T}\end{equation}
The above equation is of profound importance in this model.  This
is because,  it ensures that the equilibrium of the ejecting baryon
-radiation plasma is not disturbed, even when it flows out of concentrated
regions of cold dark matter.  This we can write, because the distribution
function of pressure of the baryon-radiation plasma depends only on
$\frac{E}{T}$.  We assume here that the chemical potential is much smaller
than the temperature.  So from small perturbations about the equilibrium
values of variables $p_{b\gamma}$ and$\varepsilon_{b\gamma}$ in the
(eq. 27), about a fixed temperature T,  we get, 
   \begin{equation} \frac{\delta p_{b\gamma}}{\delta T} = \frac{4\delta\varepsilon_{\gamma}}{9c_{s}^{2}T} 
   \end{equation}
   Thus we write:
   \begin{equation}
   \delta\varepsilon_{b\gamma} = \frac{4\delta\varepsilon_{\gamma}}{9c_{s}^{2}}\lbrace1 - \frac{\delta T}{T}\rbrace \end{equation}
   Here $\frac{\delta T}{T}\equiv\Theta$, which represents the
   anisotropy in temperature of radiation for small baryon densities
   after equality.  For small perturbations in cold dark matter, using
   (eq. 29) in (eq. 24), we write:
  
 \begin{eqnarray}
 \frac{4}{9c_{s}^{2}}\frac{\partial^{2}\delta\varepsilon_{\gamma}\lbrace 1 - \Theta\rbrace}{\partial t^{2}} - \frac{4}{9}\nabla^{2}\lbrace\delta\varepsilon_{\gamma}\lbrace
1 -\Theta\rbrace\rbrace +\nabla^{2}\phi_{o}\frac{4}
{9c_{s}^{2}}\delta\varepsilon_{\gamma}\lbrace 1 -\Theta\rbrace +\frac{4}{9c_{s}^{2}}\vec{\nabla}\cdot \lbrace\delta\varepsilon_{\gamma}(1 - \Theta)\rbrace\nabla\phi_{o} \nonumber\\
-\varepsilon_{do}\frac{\partial(\vec{\nabla}\cdot \delta\vec{v_{d}})}{\partial t}+\vec{\nabla}\cdot \lbrace\varepsilon_{b\gamma o}\frac{\partial\delta\vec{u} }{\partial t}\rbrace +\frac{4}{9c_{s}^{2}}\vec{\nabla}\cdot \lbrace\delta\varepsilon_{\gamma}(1 -\Theta)\frac{\partial\vec{u_{o}}}{\partial t}\rbrace =\sigma\nabla^{2}\delta{S(x)}
\end{eqnarray} 

For small perturbations in Cold dark matter,  we neglect the term $\vec{\nabla}\cdot \delta\vec{v_{d}} $.  This is due to cold dark matter being highly non -relativistic.  Also we neglect $ \vec{\nabla}\cdot \frac{\partial\delta\vec{u}}{\partial t} $.  This is due to the fact, that in this model,  we assume that the ejecting velocity of baryon - radiation plasma does not suffer interactions and collisions.  This is because of dark matter being truly collisionless.  So, we further write: \begin{eqnarray}
\frac{4}{9c_{s}^{2}}\frac{\partial^{2}\delta\varepsilon_{\gamma}\lbrace 1 - \Theta\rbrace}{\partial t^{2}} - \frac{4}{9}\nabla^{2}\lbrace\delta\varepsilon_{\gamma}\lbrace
1 - \Theta\rbrace\rbrace +
 4\pi G\varepsilon_{do}\frac{4}{9c_{s}^{2}}\delta\varepsilon_{\gamma}\lbrace 1 - \Theta\rbrace \nonumber \\ +\frac{4}{9c_{s}^{2}}\vec{\nabla}\cdot \lbrace\delta\varepsilon_{\gamma}(1 -\Theta)\rbrace\nabla\phi_{o} +\frac{4}{9c_{s}^{2}}\vec{\nabla}\cdot \lbrace\delta\varepsilon_{\gamma}(1-\Theta)\frac{\partial\vec{u_{o}}}{\partial t}\rbrace
 =\sigma\nabla^{2}\delta{S(x)}
 \end{eqnarray}
 In the above equation,  we  see the contribution of divergence in the
 anisotropy of radiation at epochs between equality and recombination.  At
 epochs near recombination, the strongly coupled baryon-radiation plasma
 starts rapidly to decouple from the cold dark matter.  During these
 epochs, therefore the actual velocity of the baryon -radiation plasma
 does not change appreciably due to the gravitational force of cold dark
 matter.  During these epochs, the rate of increase of ejection velocity
 is only due to gravitation potential of cold dark matter.  There is no
 force due to the pressure of the baryon -radiation plasma.  Therefore
 we write:\begin{equation}\frac{\partial\vec{u_{o}}}{\partial t} =
 -\nabla\phi_{o}\cdot \end{equation}
 We can argue that due to high baryon densities, the pressure of
 the baryon-radiation plasma vanishes at these epochs.  But the baryon
 densities can never be very high.  This is because it will hinder hydrogen
 nucleosynthesis after recombination.  Thus it has to be assumed  that
 the pressure of the  baryon - radiation plasma   vanishes not due to
 very high baryon densities, but due to a strange form of energy(dark
 energy)with negative pressure, which has started to dominate near
 recombination.  Therefore, the acceleration in the ejection velocity will
 be primarily due to the gravitational forces So, we write:
 \begin{equation}\vec{u_{o}}= \vec{v_{b\gamma o}} -\vec{v_{do}}\end{equation}
  \begin{equation} \dot{\vec{u_{o}}} = \dot{\vec{v_{b\gamma o}}} - \dot{\vec{v_{do}}}\end{equation}
 \begin{equation} \dot{\vec{v_{b\gamma o}}}= 0
 \end{equation}
 The above (eq. 35) shows that the actual velocity of the baryon-radiation
 plasma stops to increase at epochs, where it is almost about to decouple
 from cold dark matter.  Therefore, at these epochs the acceleration of cold
 dark matter reverses its sign and we write:
 \begin{equation}
 \dot{\vec{v_{do}}}=\nabla\phi_{o}
 \end{equation}
 With these Assumptions  the (eq. 31) reduces to:
 \begin{equation}\frac{\partial^{2}}{\partial t^{2}}\lbrace\delta\varepsilon_{\gamma}
 (1 - \Theta)\rbrace -c_{s}^{2}\nabla^{2}\lbrace\delta\varepsilon_{\gamma}(1 - \Theta)\rbrace +4\pi G\varepsilon_{do}\lbrace\delta\varepsilon_{\gamma}(1 - \Theta)\rbrace
 =\frac{9c_{s}^{2}}{4}\sigma\nabla^{2}\delta{S(x)}
\end{equation}
Considering adiabatic perturbations:
\begin{equation}\delta S = 0\end{equation}
We use:
\begin{equation}\delta\varepsilon_{\gamma}(1 - \Theta)(\vec{x}, t) = \int\delta\varepsilon_{\gamma k}(t)(1 - \Theta)_{k}(t)e^{i\vec{k}\cdot \vec{x}} \frac{d^{3}k}{(2\pi)^{3/2}}\end{equation} 
Then  we write:
 \begin{equation}
 \delta\varepsilon_{\gamma k}(t)(1 - \Theta)_{k}(t) = y_{k}(t)
 \end{equation}
The (eq. 37) now reduces to:
\begin{equation}
\ddot{y}_{k}(t) + k^{2}c_{s}^{2}y_{k}(t) + 4\pi G\varepsilon_{do}y_{k}(t)=0\end{equation}
\begin{center}or\end{center}
\begin{equation}
\ddot{y}_{k}(t) +(k^{2}c_{s}^{2} + 4\pi G\varepsilon_{do})y_{k}(t)=0
\end{equation}
The above equation has two solutions:$y_{k}\propto e^{\pm i\omega(t)}$, where
\begin{equation}
\omega(k)=\sqrt{k^{2}c_{s}^{2} + 4\pi G\varepsilon_{do}}=\sqrt{k^{2}c_{s}^{2} - 4\pi G\varepsilon_{b\gamma o}(1 - \frac{\varepsilon_{o}}{\varepsilon_{b\gamma o}})}
\end{equation}
Here the Jeans length is :
\begin{equation}
\lambda_{J} =\frac{2\pi}{k_{J}} = c_{s}\sqrt{\frac{\pi}
{G\varepsilon_{b\gamma o}}\lbrace 1 -
\frac{\varepsilon_{o}}{\varepsilon_{b\gamma o}}\rbrace}
\end{equation}
Thus $\omega(k)$ is real for $\lambda < \lambda_{J}$ or $k>k_{J}$, where \begin{equation}
k_{J} = \frac{2\pi}{c_{s}}\sqrt{\frac{G\varepsilon_{b\gamma o}\lbrace 1 - \frac{\varepsilon_{o}}{\varepsilon_{b\gamma o}}\rbrace}{\pi}}
\end{equation}
when 
\begin{equation} k^{2}c_{s}^{2} > 4\pi G\varepsilon_{b\gamma o}\lbrace 1 - \frac{\varepsilon_{o}}{\varepsilon_{b\gamma o}}\rbrace
\end{equation}
Where $\varepsilon_{o}$ is the total equilibrium value of energy density and $\varepsilon_{b\gamma o}$ is the equilibrium energy density of baryon-radiation plasma, at these epochs.  We interpret from above,  that all the modes near recombination are real.  This is because the value of $\frac{\varepsilon_{o}}{\varepsilon_{b\gamma o}}$ starts increasing rapidly  near recombination.  This is due to the decoupling of baryon - radiation plasma from cold dark matter.  This explains the maximum frequency of fluctuations in the CMB temperature anisotropy at red -shifts of recombination.  For $ \lambda < \lambda_{J}$ the solutions are:
\begin{equation}
y_{k}(\vec{x}, t)\propto \sin(\omega t + kx +\alpha)
\end{equation}
For $\lambda < \tau_{\gamma} < \lambda_{J}$, where $\tau_{\gamma}$
stands for the mean free path of free- streaming photons after
equality, free-streaming becomes important.  Free-streaming refers
to the propagation of photons without scattering.  We write the
equation for free-streaming photons in the conformal Newtonian gauge
\cite{Scott2003}.  The gravitational potential is primarily due to the
cold dark matter.  We argue that the photons  can still be described by
the equilibrium distribution functions.  This is because the cold Dark
matter is collisionless.  Therefore,  their mutual interactions will not
disturb the equilibrium of the ejecting baryon-radiation plasma.  Also, we
treat the baryon-radiation plasma as free-streaming photons for low
baryon-densities.  We use the metric below:
\begin{equation}
 ds^{2}= -1 -2\psi(\vec{x}, t)dt^{2} + a^{2}\delta_{ij}(1 + 2\phi(\vec{x}, t))dx^{i}dx^{j}
 \end{equation}
where $\psi$ corresponds to the Newtonian potential and $\phi$ is the
perturbation to the spatial curvature. 
The equation for free- streaming photons is:
\begin{equation}
\frac{1}{p}\frac{dp}{dt}= -H -\frac{\partial\phi}{\partial t} - \frac{\hat{p}^{i}}{a}\frac{\partial \psi}{\partial x^{i}}
\end{equation}
\begin{equation}
\psi =\phi
\end{equation}
We can make the above assumption, for epochs between equality and
recombination.  This is because after Equality, and before recombination,  the
dominant contribution to the potential is due to cold dark matter.  So, we
neglect the non-diagonal components of the energy-momentum tensor of
the baryon-radiation plasma.  Recall that the difference $\psi - \phi$ is
very small compared to $\psi$.  Thus we assume it to be a single
perfect fluid.  In a static Einstein universe, we neglect the terms due
to the Hubble parameter $H$.  At scales of the order of mean-free path
of the free-streaming photons in the baryon-radiation plasma, i. e.  of
the order of $\tau_{\gamma}$, we neglect the spatial inhomogeneities
$\frac{\partial\phi}{\partial\vec{x}}$.  We therefore write:
\begin{equation}
\frac{1}{p}\frac{dp}{dt} = -\frac{\partial\phi}{\partial t}
\end{equation} 
Using the assumption of (eq. 50), we write:
\begin{equation}
\frac{1}{p}\frac{dp}{dt} =-\frac{\partial\psi}{\partial t}
\end{equation}
Recall that the Newtonian potential $\psi$  corresponds to the dark
matter potential $\phi_{do}$.  This $\phi_{do}$ remains constant between
epochs of equality and recombination.  Thus we write:
\begin{equation}
\log p = -\psi \end{equation}
 \begin{equation}\log p = -\phi_{do}\Rightarrow p=e^{-\phi_{do}}\Rightarrow\delta p=-e^{-\phi_{do}}\delta\phi_{d}\end{equation} 
The momentum per unit volume of radiation at epochs of decoupling is
equivalent to radiation pressure.  Also, if the photons,  just at the epochs
of decoupling are in thermal equilibrium, and that it has same average
energy associated with each independent degree of freedom, we write:
\begin{equation}
\delta p_{\gamma}=\frac{1}{3}\delta p
\end{equation}
where $\delta p_{\gamma}$ is the increment in radiation pressure
associated with each independent degree of freedom at decoupling.For a unit volume ,decoupling photons' energy density is same as radiation
pressure per degree of freedom, we therefore write:
\begin{equation}
\delta\varepsilon_{\gamma}= -\frac{1}{3}e^{-\phi_{do}}\delta\phi_{d}
\end{equation}
We use (eq. 40 and (eq. 47) to write:
\begin{equation}\delta\varepsilon_{\gamma k}(1 - \Theta)(\vec{x}, t) = A\sin(\omega t + kx + \alpha)\end{equation} 
where \begin{equation}\omega = \sqrt{k^{2}c_{s}^{2} - 4\pi G\varepsilon_{b\gamma o}(1 - \frac{\varepsilon_{o}}{\varepsilon_{b\gamma o}})}\end{equation}
Therefore,  for epochs of decoupling, we write: \begin{equation}
\frac{-e^{-\phi_{do}}\delta\phi_{d}}{3}\lbrace 1 - \Theta\rbrace  = A\sin(\omega t + kx + \alpha) \end{equation} \begin{center}or\end{center}
\begin{equation}
\lbrace\Theta -1\rbrace = \frac{3e^{\phi_{do}}}{\delta\phi_{d}}A\sin(\omega t + kx + \alpha) \end{equation}
 From the above equation we can write:
 \begin{equation}
 \frac{\delta\phi_{d}}{3}=\lbrace\theta
+\phi_{d}(1+\theta)^{2}\rbrace
\end{equation}  
  Therefore, we write:
  \begin{equation}
  \theta \approx\frac{\delta\phi_{d}}{3}
  \end{equation}
 The above result is the same as that predicted by Sachs-Wolfe.Here it is important to note that in the (eq.60),the problem of divergences in the temperature anisotropy  does not arise as the term $\delta\phi_{d}$ represents the average perturbation in the gravitational potential of the cold dark matter between the epochs of equality and recombination.We know this can't take zero value. Three
 types of effects(due to fluctuations in density, velocities
 and potential)simultaneously contribute to the CMB temperature
 anisotropy.  The fluctuations that matter at scales beyond $1^{\circ}$
 are those in the gravitational potential $\delta\phi_{d}$(Sachs-Wolfe
 effect)\cite{Sach1967}.\\
 \indent 
The above equation represents the anisotropy in the CMB temperature, for
epochs near recombination, for regions where sufficient primordial helium
synthesis takes place, even before recombination, or the dark energy has
started to dominate.  Recall that in assumption of(eq. 32), we argue that
the force due to pressure of baryon-radiation plasma vanishes near
recombination epochs, and that it can be only due to the fact that the
dark energy with negative pressure has started to dominate.  The value of
$\phi_{do}$ is constant.  This is because the potential of cold dark matter
remains constant for many epochs between equality and recombination.  Thus
in a static universe, with this model,  we see that the dominant component
in CMB temperature anisotropy fluctuations is near recombination.  This
is because,  at epochs near recombination, the perturbations in the cold
dark matter potential are very small.  This is because in this model, the
decoupling of the baryon - radiation plasma from concentrated regions
of cold dark matter, near epochs of recombination is almost near
completion. \\
\indent
We now write the CMB fluctuations for supercurvature modes i. e modes with
$\lambda >> \lambda_{J}$, and for regions where sufficient primordial
helium synthesis takes place, even before recombination, i. e.  the
baryon-densities are high.  We neglect the effect of gravity at these
epochs.  We can do so because at late recombination epochs, the pressure of
baryon-radiation plasma is low, and the negative pressure of a strange form
of energy(dark energy), which starts to dominate at these epochs,  cancels
the effect of forces due to low pressure baryon-radiation plasma and
that of gravity. we therefore write:\begin{equation}y_{k}\propto e^{\pm
ikc_{s}t}\end{equation}
Also for $\lambda >>\lambda_{J}$, free-streaming of photons is no longer
relevant.  Therefore the scattering of photons will dilute the anisotropy
to a large extent. 
For late      recombination epochs, when the radiation has almost decoupled
from matter,  we write:
  \begin{equation}c_{s}^{2} = \frac{1}{3}\Rightarrow y_{k}\propto e^{\pm \frac{ikt}{\sqrt{3}}}\end{equation}

We have neglected gravity here again, because at these epochs, the effect
of dark energy,  which had started to dominate from some earlier epochs is
to cancel the forces due to gravity and the pressure of baryon-radiation
plasma.  This is possible even in regions of lower baryon densities, which
has higher pressure than the regions of higher baryon densities.  This is
only because the dark energy with negative pressure had been dominating
from some earlier recombination epochs.  The above equation shows that
the frequency of fluctuations in CMB temperature anisotropy spectrum
of supercurvature modes(modes which enter the horizon very early
near recombination epochs),  at late recombination,  remain constant
till today.  This is valid both for regions of low or higher baryon
densities(where sufficient primordial helium nucleosynthesis takes place
before recombination).  This is because at very late recombination epochs
when the radiation has almost decoupled fully from matter, the speed of
sound approaches a constant value of $\frac{1}{3}$. 
The effect of scattering of photons will dilute this anisotropy in
supercurvature modes.  So, it is of not much cosmological significance. 

\section{Instability in Expanding Universe: Newtonian theory}
\label{sec:3}

Using our model (see Sec. ~\ref{sec:1 Intro}), we study the same
scenario, i. e.  of cold dark matter in the presence of strongly coupled
baryon -radiation plasma at epochs between equality and recombination.  Here
we use Newtonian theory of expanding universe \cite{Bonnor1957}.  We
treat the dynamics between cold dark matter and the baryon - radiation
plasma,  again in the framework of the Merscerskii equation.We ensure the collisionless nature of cold dark matter by the same inherent assumption as in (sec.~\ref{sec:2}).We assume
that the strongly coupled baryon -radiation plasma is in the presence
of a gravitational potential.  This potential,  is only due to cold dark
matter, at epochs after equality and before recombination.  We neglect
the non-diagonal components of the energy-momentum tensor of the cold
dark matter.  This is because, the difference $\psi - \phi$ is very small
as compared to $\psi$.  We consider dark matter as highly non -relativistic
fluid compared to the baryon - radiation plasma. We consider epochs when
the Dark matter has already started clustering.  We neglect the rate of
flow of cold dark matter out of a given region of space, as compared to
the baryon-radiation plasma.  Therefore we write:
\begin{equation}
  \frac{\partial\varepsilon_{b\gamma}}{\partial t} + \vec{\nabla}\cdot (\varepsilon_{b\gamma}\vec{u}) = 0\end{equation}
where $\vec{u}$ is the relative velocity of the baryon - radiation
plasma with respect to the cold dark matter.  We write the Merscerskii
equation for cold dark matter with baryon radiation plasma,  ejecting
out of concentrated regions of space dominated  by cold dark matter:
  \begin{equation}
    \varepsilon_{do}\lbrace\frac{\partial\vec{v_{do}}}{\partial t}  + (\vec{v_{do}}\cdot \vec{\nabla})\vec{v_{do}}\rbrace = -\varepsilon_{do}\vec{\nabla}\phi_{o} -\vec{\nabla} p_{b\gamma} -\frac{\partial\varepsilon_{b\gamma o}}{\partial t}\vec{u_{o}} \end{equation} In an expanding flat, Isotropic and homogenous universe, we write:
 \begin{equation} \vec{v_{b\gamma}}=\vec{v_{b\gamma o}}(t); \varepsilon_{b\gamma}= \varepsilon_{b\gamma o}(t); 
 \vec{u_{o}}= H(t). \vec{x} ;  \varepsilon_{do} =\varepsilon_{do}(t);
\vec{v_{do}}=\vec{v_{do}(t)}\end{equation}
 Therefore we write:
 \begin{equation}
 \frac{\partial\varepsilon_{b\gamma o}(t)}{\partial t} + \varepsilon_{b\gamma o}(t)\vec{\nabla}\cdot \vec{u_{o}} =0 
 \end{equation}
We first discuss for scales $>>H^{-1}$ i. e the curvature scale.  In this
case,  we neglect the velocities of Cold dark matter particles.  This is
because,  there is not enough time to move highly non -relativistic cold
dark matter upto distances greater than the Hubble scale.  Therefore the
entropy per cold dark matter particle remains constant on supercurvature scales($(k\eta
> 1)$).  So we write:
 \begin{equation}
 \frac{\partial\varepsilon_{b\gamma o}(t)}{\partial t} + 3\varepsilon_{b\gamma o}(t)H(t) =0 
 \end{equation}
The (eq. 66) gives: 
 \begin{equation}
  \varepsilon_{do}\vec{\nabla}\phi_{o} + \vec{\nabla} p_{b\gamma} + \frac{\partial\varepsilon_{b\gamma o}}{\partial t}\vec{u_{o}} = 0\end{equation}
 We take the divergence of the above equation  to write:
  \begin{equation}
  6H^{2}\varepsilon_{b\gamma o} = 4\pi G\varepsilon^{2}_{do}
  \end{equation}
  To get the above equation, for near recombination epochs,  we use:
  \begin{equation}
  \nabla^{2}p_{b\gamma}=0\end{equation}
We use the above assumption because for epochs near recombination, the
baryon-radiation plasma starts to decouple fast.  Therefore, only the
gravitational force determines the acceleration of dark matter.  Now for
Adiabatic perturbations, we write:
  \begin{equation}
  \delta S = 0
  \end{equation}
  Also,  we write the following perturbations for other variables:
  \begin{equation}
  \varepsilon_{d} = \varepsilon_{d} = \varepsilon_{do} + \delta\varepsilon_{d}(\vec{x}, t);
  \vec{u} = \vec{u_{o}} + \delta\vec{u}(\vec{x}, t);\phi = \phi_{o} + \delta\phi ;
  p_{b\gamma} = p_{b\gamma o} + \delta p_{b\gamma} = p_{o} + c_{s}^{2}\delta\varepsilon_{b\gamma}
  \end{equation}
Where the variables have their usual meanings (see Sec. ~\ref{sec:2}).  We
use (eq. 71) in (eq. 66) and (eq. 68) to write :
  \begin{equation}
  \frac{\partial\delta\varepsilon_{b\gamma}}{\partial t} = \varepsilon_{b\gamma o}\vec{\nabla}\cdot \delta\vec{u}  + \vec{\nabla}\cdot \lbrace\delta\varepsilon_{b\gamma}\vec{u_{o}}\rbrace = 0
  \end{equation}
 \begin{center} and \end{center}
  \begin{equation}
  \varepsilon_{do}\vec{\nabla}\delta\phi
  + \delta\varepsilon_{d}\vec{\nabla}\phi_{o} + \vec{\nabla}\delta p_{b\gamma o}+\frac{\partial\varepsilon_{b\gamma o}}{\partial t}\delta\vec{u}+ \frac{\partial\delta\varepsilon_{b\gamma}}{\partial t}\vec{u_{o}} = 0
  \end{equation} 
  We use the Langragian co-ordinates and write:
  \begin{equation}
  \lbrace\frac{\partial}{\partial t}\rbrace_{x} =\lbrace\frac{\partial}{\partial t}\rbrace_{q} -\vec{u_{o}}\cdot \vec{\nabla}_{x}
  \end{equation}
    where \begin{equation}
    \vec{u_{o}} = H(t)q  ; x =aq
    \end{equation}
    We write:\begin{equation}\vec{\nabla}_{x} = \frac{1}{a}\vec{\nabla}_{q} ;  \delta = \frac{\delta\varepsilon}{\varepsilon_{o}} 
   \end{equation}
where $\delta$ is the fractional amplitude of  perturbations.  We write
(eq. 75) in the Co-moving coordinates, using (eq. 65) and (eq. 77) to get:
   \begin{equation}
   \dot{\delta_{b\gamma}} + \frac{\vec{\nabla}\cdot \delta\vec{u}}{a} = 0
   \end{equation}
  also we write the (eq. 76) in the co-moving coordinates to get:
  \begin{equation}
  \frac{\varepsilon_{do}}{a}\vec{\nabla}\delta\phi + \frac{\varepsilon_{do}}{a}\delta_{d}\vec{\nabla}\phi_{o} + \frac{c_{s}^{2}}{a}\varepsilon_{b\gamma o}\vec{\nabla}\delta_{b\gamma}+\dot{\varepsilon_{b\gamma o}}\delta\vec{u}  +\dot{\varepsilon_{b\gamma o}}\delta_{b\gamma}+\varepsilon_{b\gamma o}\dot{\delta_{b\gamma}} -\frac{\vec{u_{o}^{2}}}{a}\varepsilon_{b\gamma o}\vec{\nabla}\cdot \delta_{b\gamma} = 0
   \end{equation}
We write (eq. 69)in co -moving coordinates:
   \begin{equation}
   \dot{\varepsilon_{b\gamma o}} = -\frac{3\varepsilon_{b\gamma o}(t)H(t)}{a} 
   \end{equation}
 from (eq. 71),  we get:
   \begin{equation}
   \varepsilon_{do}\vec{\nabla}. \delta_{d} = \frac{6H^{2}\varepsilon^{2}_{b\gamma o}\vec{\nabla}. \delta_{b\gamma}}{8\pi G\varepsilon_{do}}
   \end{equation}
 We take the divergence of (eq. 76), and use (eqn's. 68-77-79) and (eq. 80) to get:
  \begin{equation}
  \frac{12H^{2}\delta_{d}}{a}+ 3aH\dot{\delta_{b\gamma}}+ \lbrace\frac{c_{s}^{2}-u_{o}^{2}}{a}\rbrace\nabla^{2}\delta_{b\gamma} + \vec{\nabla}. \dot{\delta_{b\gamma}} -\vec{\nabla}. \delta_{b\gamma}\lbrace\frac{H(2u_{o} + 3)}{a}\rbrace = 0
  \end{equation}
We Neglect the fourth term i. e. $\vec{\nabla}. \dot{\delta_{b\gamma}}$. This is
because, it is proportional to $\frac{\vec{\nabla}. \vec{\nabla}\delta u}{a}$, which is
very small. This is due to very small divergence in the perturbations
to  the ejection velocity of baryon -radiation plasma. With this
assumption, again we argue that the cold dark matter is truly
collisionless. So, in this model, it does not disturb the equilibrium of
the ejecting baryon-radiation plasma. From (eq. 71), we get:
  \begin{equation}8\pi G \varepsilon^{2}_{do}\delta_{d}=6H^{2}\varepsilon^{2}_{b\gamma o}\delta_{b\gamma}
  \end{equation} 
Using the above, we write (eq. 84) as:
  \begin{equation}
   \dot{\delta_{b\gamma}} +\frac{3H^{3}\varepsilon^{2}_{b\gamma}\delta_{b\gamma}}{8\pi Ga^{2}\varepsilon^{2}_{do}}+ \lbrace\frac{c_{s}^{2}- u^{2}_{o}}{3a^{3}H}\rbrace\nabla^{2}\delta_{b\gamma} + \frac{\vec{\nabla}. \delta_{b\gamma}H^{2}}{3a^{2}}\lbrace2u_{o}+ 3\rbrace = 0  
   \end{equation}
   Therefore, neglecting the term containing $\vec{\nabla}. \delta_{b\gamma}$, we write:
   \begin{equation}
   \dot{\delta_{b\gamma}} + \frac{3H^{3}\varepsilon^{2}_{b\gamma}\delta_{b\gamma}}{8\pi Ga^{2}\varepsilon^{2}_{do}}+\lbrace\frac{c_{s}^{2} - u^{2}_{o}}{3a^{3}H}\rbrace\nabla^{2}\delta_{b\gamma} = 0
   \end{equation}
 We use: \begin{equation}\delta_{b\gamma}= \int\delta_{b\gamma k}(t)e^{i\vec{k}\cdot \vec{q}}\frac{d^{3}k}{\sqrt{(2\pi)^{3}}}   
 \end{equation}
 Thus,  we write the equation for the evolution  of fractional amplitudes
 of perturbations in the baryon -radiation plasma as:
 \begin{equation}
 \delta_{b\gamma} = \exp\lbrace k^{2}(\frac{c_{s}^{2} - u^{2}_{o}}{3a^{3}H}) -\frac{3H^{3}\varepsilon^{2}_{b\gamma o}}{8\pi Ga^{2}\varepsilon^{2}_{do}}\rbrace t \end{equation} 

For very late  recombination epochs, the radiation starts to decouple
rapidly from matter. Therefore, baryon densities are very low. So we write:
 \begin{equation}
 c^{2}_{s}=\frac{1}{3}
 \end{equation}
 where $c^{2}_{s}$ is the speed of sound. 

The (eq. 89) shows,  that for low baryon densities,   fractional amplitudes
of perturbations in the radiation,  which originate at late recombination
epochs grow at the  fastest rate.  This is because the value of $c_{s}^{2}$
is maximum at late recombination epochs. It is because of the lowest baryon
densities, in the coupled baryon -radiation plasma at these epochs. Also
we can see that the second term in the exponent in (eq. 89)vanishes for
very late recombination epochs, as ratio $\frac{\varepsilon^{2}_{b\gamma
o}}{\varepsilon^{2}_{do}}\rightarrow 0$. We thus write the equation for
evolution of the fractional amplitude of perturbations in the Cold dark
matter, which originate after equality,  when the clustering of dark matter
had already started. 
 \begin{equation}
 \delta_{d} = \frac{3H^{2}}{4\pi G}\frac{\varepsilon^{2}_{b\gamma o}}{\varepsilon^{2}_{do}}\exp \lbrace k^{2}(\frac{c_{s}^{2} - u^{2}_{o}}{3a^{3}H}) -\frac{3H^{3}\varepsilon^{2}_{b\gamma o}}{8\pi Ga^{2}\varepsilon^{2}_{do}}\rbrace t
 \end{equation}
The above equation represents the growth of the fractional amplitudes of perturbations in the cold dark matter,  which originate after equality, for scales $>>H^{-1}$.It is important to note that for late recombination epochs,it reduces to the standard result $\delta_{d}\propto a$.
 If baryons contribute  a significant factor of the total matter
density, CDM growth rate will be slowed down between equality and the
recombination epochs \cite{Hu1996}. Also we see that the perturbation
growth rate will slow with scale factor \cite{Caldwell2002}. The
CDM density fluctuations will dominate the density perturbations
of baryon-radiation plasma \cite{Hu1996}. This is because for scales
$>>H^{-1}$, the density perturbations of baryon-radiation plasma are washed
out by the scattering of photons at scales $>\tau_{\gamma}$, which is the
mean free path of photons. The perturbations in the cold dark matter will
cease to grow when the dark energy starts to dominate \cite{Turner2008}. We
can interpret this from (eq. 91). This is because with the growth of dark
energy, the value of $c_{s}^{2}$ will decrease. This will then lead to
ceasing of growth of CDM perturbations. There is existence of non-linear
structures today. This implies that the growth of fluctuations must have
been driven by non-baryonic dark matter, which was not relativistic at
recombination.  Also,  we see that the perturbations  at supercurvature
scales grow slowly. Recall that these are the modes which enter the
horizon very early, well before the recombination epochs. The slow growth
of such modes is because, it is only at late recombination epochs
that the second term in the exponent in (eq. 91) will vanish due to
$\frac{\varepsilon_{b\gamma o}}{\varepsilon_{do}}\rightarrow 0$. Also, it is
only at late recombination epochs, that the value of $c_{s}^{2}$ reaches
its maximum value of $\frac{1}{3}$, just before decoupling. The amplitude
of the fractional density perturbations in the cold dark matter, in
(eq. 91) will be maximum when the ratio$\frac{\varepsilon_{b\gamma
o}}{\varepsilon_{do}}\rightarrow 1$. This will occur when
 \begin{equation}
  \varepsilon_{do}=\frac{3H^{2}}{2\pi G}
  \end{equation}
  In writing the above equation, we use the result of (eq. 71). So for epochs when the density of baryon-radiation plasma is equal to the density of the cold dark matter, (eq. 91) is:
 \begin{equation}
 \delta_{d} = \frac{\varepsilon_{do}}{2}\exp \lbrace k^{2}(\frac{c_{s}^{2} - u^{2}_{o}}{3a^{3}H}) -\frac{H\varepsilon_{do}}{4a^{2}}\rbrace t
 \end{equation} 

Now we discuss the gravitational instability for scales $<H^{-1}$. This
originates after equality,with dark matter dominance in the presence of
strongly coupled baryon - radiation plasma in an expanding universe. We
discuss it in the Newtonian theory. The decoupling of the strongly
coupled baryon - radiation plasma from the non -relativistic cold
dark matter starts after equality. Let us assume that the separation
of this plasma from the cold dark matter gives a relative velocity of
$\vec{u_{o}}$ to the baryon - radiation plasma. At small scales with low
baryon densities, we neglect the contribution of non-diagonal components
in the energy- momentum tensor of dark matter. Therefore, we treat it as
a perfect fluid for many epochs between equality and recombination. We
treat the strongly coupled baryon - radiation plasma as a perfect fluid
for epochs between equality and recombination. This is because of low
baryon -densities at these epochs. Recall that the baryon densities only
starts increasing substantially after recombination. This is when the
primordial nucleosynthesis of hydrogen and helium will start. However, as
an exception in certain regions, the primordial nucleosynthesis may start
at epochs before recombination. We now conclude that there is sufficient
time for the Cold dark matter to flow through distances at scales $<
H^{-1}$. This is because from (eq. 91), we see that the perturbations in the
cold dark matter,  at small scales,  grows at the fastest rate. Therefore
we write the Merscerskii equation as below:
 \begin{equation}
 \varepsilon_{do}\lbrace\frac{\partial\vec{v_{do}}}{\partial t} + (\vec{v_{do}}\cdot \vec{\nabla})\vec{v_{do}}\rbrace = -\varepsilon_{do}\vec{\nabla}\phi_{o}
-\vec{\nabla} p_{b\gamma}-\frac{\partial\varepsilon_{b\gamma o}}{\partial t}\vec{u_{o}}   
\end{equation}
where \begin{equation}\vec{}u_{o}=\vec{v_{b\gamma o}}- \vec{v_{do}}\end{equation}

At small scales, we assume that in the time that the baryon -radiation
plasma flows out of a given region of space, the inhomogeneity in Cold
dark matter in that time duration is negligible. So we assume that the
baryon-radiation plasma can flow out of a concentrated region of dark
matter without generating the collision terms. This is because the cold
dark matter is highly non-relativistic and collisionless. Also, because
the dark matter has already started to cluster. Therefore we write:
\begin{equation}\vec{v_{b\gamma}}=\vec{v_{b\gamma o}(t)} = H(t)\vec{x}\end{equation}
where $\vec{x}$ is the eulerian cordinate. We write the continuity equation
for flow of the strongly coupled baryon - radiation plasma as :
\begin{equation}
\frac{\partial\varepsilon_{b\gamma}}{\partial t} +\vec{\nabla}. (\varepsilon_{b\gamma}\vec{v_{b\gamma}})= 0
\end{equation}
Here $\varepsilon_{b\gamma}$ is equal to $\varepsilon_{b\gamma o}(t)$
\begin{equation}
\dot{\varepsilon_{b\gamma o}} = -3H\varepsilon_{b\gamma o}
\end{equation}
We take the divergence of (eq. 94) and neglect the spatial dependence of
$\vec{u_{o}}$.  This is due to the collisionless nature of dark matter. We
also assume that $\varepsilon_{do}$ is constant for small scales. We thus
write the Friedmann equation:
\begin{equation}
\dot{H} + H^{2} = -\frac{4\pi G\varepsilon_{do}}{3}
\end{equation}
For adiabatic perturbations we write from :
   \begin{eqnarray}
\varepsilon_{d}(\vec{x}, t)=\varepsilon_{do} + \delta\varepsilon_{d}(\vec{x}, t);
\vec{v_{d}}(\vec{x}, t)= \vec{v_{do}} + \delta\vec{v_{d}}(\vec{x}, t)= \delta\vec{v_{d}}(\vec{x}, t) ;
\phi(\vec{x}, t)= \phi_{o} +  \delta\phi (\vec{x}, t);\nonumber \\
\vec{u}(\vec{x}, t)= \vec{u_{o}} + \delta\vec{u}(\vec{x}, t);
p_{b\gamma}=p_{b\gamma o}+ c^{2}_{s}\delta\varepsilon_{b\gamma};\delta S=0\end{eqnarray}
where the variables have their usual meanings (see Sec. ~\ref{sec:2}). Then we use the perturbed values of variables in (eq. 94),  to write:
  \begin{eqnarray}
\varepsilon_{do}\delta\dot{\vec{v_{b\gamma}}}-\varepsilon_{do}\dot{\delta\vec{u}}+\delta\varepsilon_{d}\dot{\vec{v_{b\gamma o}}}-\delta\varepsilon_{d}\dot{\vec{u_{o}}} + \varepsilon_{do}(\vec{v_{b\gamma o}}\cdot \vec{\nabla})\delta\vec{v_{b\gamma}} -\varepsilon_{do}\vec{u_{o}}\cdot \vec{\nabla}\delta\vec{v_{b\gamma}} - \varepsilon_{do}\vec{v_{b\gamma o}}\cdot \vec{\nabla}\delta\vec{u}\nonumber\\
+ \varepsilon_{do}\vec{u_{o}}\cdot \vec{\nabla}\delta\vec{u}=-\varepsilon_{do}\vec{\nabla}\delta\phi -\delta\varepsilon_{d}\vec{\nabla}\phi_{o}-c_{s}^{2}\vec{\nabla}\varepsilon_{b\gamma}-\dot{\varepsilon_{b\gamma o}}\delta\vec{u} -\delta\dot{\varepsilon_{b\gamma}}\vec{u_{o}}
\end{eqnarray}
   We then use the perturbed values of variables in (eq. 97) to get:
   \begin{equation}
   \delta\dot{\varepsilon_{b\gamma}} +\varepsilon_{b\gamma o}(t)\vec{\nabla}. \delta\vec{v_{b\gamma}}+
   \vec{v_{b\gamma o}}\vec{\nabla}. \delta\varepsilon_{b\gamma} =0
   \end{equation}
   We now write (eq. 98) in co-moving coordinates. 
   \begin{equation}
   \dot{\varepsilon}_{b\gamma o} = -\frac{3\varepsilon_{b\gamma o}H}{a}
   \end{equation}
   We also write (eq. 102) in co-moving coordinates to get:
   \begin{equation}
   \dot{\delta\varepsilon_{b\gamma}}=-\frac{\varepsilon_{b\gamma o}}{a}\vec{\nabla}. \delta\vec{v_{b\gamma}}
   \end{equation}

Using (eq. 95) in (eq. 94) and also using the (eq. 103),  (eq. 104), we write
(eq. 94) in co-moving coordinates. We then take the divergence of the
obtained equation and use the results of (eq. 112) and (eq. 117), to write:
   \begin{eqnarray}
   \ddot{\delta_{b\gamma}} +H\lbrace 1 + \frac{3}{a} -\frac{u_{o}}{aH}\rbrace\dot{\delta_{b\gamma}} -\lbrace\frac{3\dot{H}}{a}+\frac{3H^{2}\varepsilon_{b\gamma o}}{\varepsilon_{do}}+\frac{9H^{2}\varepsilon_{b\gamma o}}{a\varepsilon_{do}}-\frac{3Hu_{o}}{a^{2}}+4\pi G\varepsilon_{b\gamma o}-\frac{4\pi G\varepsilon_{b\gamma o}}{a}\rbrace\delta_{b\gamma}\nonumber\\ -\frac{3H\varepsilon_{b\gamma o}}{a^{2}\varepsilon_{do}}\lbrace av_{b\gamma o}-1-u_{o}\rbrace\vec{\nabla}. \delta_{b\gamma}
-\frac{\varepsilon_{b\gamma o}\vec{\nabla}. \dot{\delta_{b\gamma}}}{a\varepsilon_{do}}+\frac{\varepsilon_{b\gamma o}}{\varepsilon_{do}}\lbrace\frac{c_{s}^{2}-v_{b\gamma o}u_{o}}{a^{2}}\rbrace\nabla^{2}\delta_{b\gamma}=0
\end{eqnarray}
   Also we can write:
\begin{equation}\delta\varepsilon_{b\gamma}=\frac{4c_{s}^{2}\delta\varepsilon_{\gamma}(1-\Theta)}{9}\end{equation}
where $\Theta=\frac{\delta T}{T}$, represents the anisotropy in the
temperature of radiation. 
The total energy distribution composed of the sum of the dark matter
and the baryon-radiation plasma is a function of time in an expanding
universe. If this remains smooth inspite of the inhomogeneities at scales
smaller than $H^{-1}$, we can write:
\begin{equation}
\varepsilon_{b\gamma}a^{4}(t) + \varepsilon_{d}a^{3}(t)= E_{total}(t)\end{equation}
or for any time $'t'$, we can write:
\begin{equation}-a\delta\varepsilon_{b\gamma}= \delta\varepsilon_{d} \end{equation} 
At epochs near recombination, when the baryon-radiation plasma is decoupled
from the cold dark matter,  we assume that their bulk velocity does
not appreciably change in these epochs.  Therefore, it remains constant
for epochs near recombination. The ejection velocity of baryon-radiation
plasma is affected only due to the gravitational force of the dark matter
potential. There is no force due to the pressure of the baryon-radiation
plasma. This can be either due to low pressure of the plasma, due to
higher baryonic densities, or due to some new force of strange energy(dark
energy), having negative pressure. The baryon densities can never be very
high before recombination. This is because it would hinder future hydrogen
nucleosynthesis after recombination. So, it would be rightly concluded that
a new form of dark energy having negative pressure starts to dominate
at these epochs. So, we write:
 \begin{equation}
 \vec{v_{b\gamma o}}=\text{constant}\Longrightarrow\dot{\vec{v_{b\gamma}}}=0 \end{equation}
 \begin{equation}
 \dot{\vec{v_{do}}}=\vec{\nabla}\phi_{o}
 \end{equation}
The above equation comes from the same argument, which we give in
(eq. 36). The actual velocity of the baryon-radiation plasma stops to
increase, when it is almost about to decouple from dark matter. At these
epochs, the acceleration of dark matter will reverse its sign.  Using the
above equation and (eq. 95), we write:
 \begin{equation} \dot{\vec{u_{o}}}=-\vec{\nabla}\phi_{o}\end{equation}
  and in co-moving coordinates,  \begin{equation}
  \dot{\vec{u_{o}}}=-\frac{\vec{\nabla}\phi_{o}}{a}\end{equation}
  \begin{equation}
  \vec{\nabla}. \delta\dot{u} =-\vec{\nabla}\delta\phi\end{equation} 
  and in co-moving coordinates, 
  \begin{equation}
  \vec{\nabla}. \delta\dot{u} =-\frac{\vec{\nabla}\delta\phi}{a}
  \end{equation}
 or \begin{equation}
 \vec{\nabla}. \delta\dot{u}=-\frac{\nabla^{2}\delta\phi}{a} =-\frac{4\pi G\delta\varepsilon_{d}}{a}\end{equation}
Using (eq. 108), we can write the above equation as:
\begin{equation}
\vec{\nabla}. \delta\dot{u}=4\pi G\varepsilon_{b\gamma o}\delta_{b\gamma}\end{equation}
 Also we assume, 
 \begin{equation}\vec{\nabla}. \dot{\delta u}=0\end{equation}

For scales much smaller than the Jeans length$\lambda < \tau_{\gamma}<
\lambda_{J}$ i. e.  scales much smaller than the curvature scale i. e for
which $H^{-1}$ is dominant, we neglect the terms $\propto H$ and write
Eq. $105$ as:
 \begin{equation}
 \ddot{y_{k}}-\frac{u_{o}}{a}\dot{y_{k}}+\lbrace 4\pi G\varepsilon_{b\gamma o}(1-\frac{1}{a})-k^{2}\varepsilon_{b\gamma o}(\frac{c_{s}^{2}-v_{b\gamma o}u_{o}}{a^{2}})-\frac{\dot{u_{o}}}{a}\rbrace y_{k}=0
 \end{equation}
 We use (eq. 39)and (eq. 40) in writing the above equation. If we now choose: 
 \begin{eqnarray}
 2\lambda = -\frac{u_{o}}{a}\nonumber \\
 \lbrace 4\pi G\varepsilon_{b\gamma o}(1-\frac{1}{a})-k^{2}\varepsilon_{b\gamma o}(\frac{c_{s}^{2}-v_{b\gamma o}u_{o}}{a^{2}})-\frac{\dot{u_{o}}}{a}\rbrace =\omega^{2}
  \end{eqnarray}
 Then the auxillary equation is:
 \begin{equation}
 D^{2}+2\lambda D + \omega^{2}=0
 \end{equation}
 with
 \begin{equation}
 D=-\lambda \pm \sqrt{\lambda^{2}-\omega^{2}}
 \end{equation}
 The only solution which is of physical significance is when $\lambda <
 \omega$, i. e.  when the roots of the auxillary equation are imaginary, i. e. :
 \begin{eqnarray}
 D=-\lambda + i\alpha \nonumber \\ 
 \alpha^{2}=\omega^{2}-\lambda^{2}\nonumber \\
 y_{k}=A\sqrt{\lbrace 1 + (\frac{\lambda}{\alpha})^{2}\rbrace}e^{-\lambda t}\cos\lbrace\alpha -\arctan\frac{\lambda}{\alpha}\rbrace
 \end{eqnarray} 
 
 Thus we write:
 \begin{equation}
\delta\varepsilon_{\gamma}\lbrace 1 - \Theta\rbrace =A\sqrt{\lbrace 1 + (\frac{\lambda}{\alpha})^{2}\rbrace}e^{-\lambda t}\cos\lbrace\alpha -\arctan\frac{\lambda}{\alpha}\rbrace
 \end{equation}
for $\lambda < \tau_{\gamma}< \lambda_{J}$,  we can write for free -streaming photons:
\begin{equation}
 \frac{1}{p}\frac{dp}{dt} = -H - \frac{\partial\phi}{\partial t} - \frac{p^{i}}{a}\frac{\partial\psi}{\partial x^{i}}\end{equation}
 where $\psi$ corresponds to the Newtonian potential in Perturbed FRW
 universe. Here we assume,   \begin{equation}\phi = \psi\end{equation}
 This is because the potential is dominated by the cold dark matter
 between epochs of equality and recombination.   We neglect the spatial
 inhomogeneities in the Newtonian potential at very small scales
 and  terms $\propto H$.  Using the same argument that we give for
 (see eq.55-56 sec.~\ref{sec:2}), we get:
 \begin{equation}
 \delta\varepsilon_{\gamma}= \frac{1}{3}\delta p = -\frac{e^{-\phi_{do}}\delta\phi_{d}}{3}
 \end{equation}
 Using (eq. 123)and (eq. 126), we write:
  \begin{equation}
  \frac{-e^{-\phi_{do}}\delta\phi_{d}\lbrace 1 -\Theta\rbrace}{3} =A\sqrt{\lbrace 1 + (\frac{\lambda}{\alpha})^{2}\rbrace}e^{-\lambda t}\cos\lbrace\alpha -\arctan\frac{\lambda}{\alpha}\rbrace 
  \end{equation}
  \begin{center}or
  \end{center}
  \begin{equation}
  \lbrace\Theta -1\rbrace = \frac{3A e^{\phi_{do}}}{\delta\phi_{d}}\sqrt{\lbrace 1 + (\frac{\lambda}{\alpha})^{2}\rbrace}e^{-\lambda t}\cos\lbrace\alpha -\arctan\frac{\lambda}{\alpha}\rbrace
  \end{equation}

This equation is of same form as the one we get for Einstein's static
universe in Jeans theory. In the static universe,  we arrived at this
equation by an assumption of (eq. 32). This we reasoned,   may be partly
possible where baryon densities are higher due to early primordial
synthesis of Helium before recombination. But the primary reason would
be that dark energy, with negative pressure starts to dominate from
these epochs. The similarity in form of the two equations,  shows that
the epochs, for which the  CMB temperature anisotropy derived from the
solutions in the static Einstein universe and those from the expanding
Newtonian match,  will be the epochs when the dark energy starts to
dominate.This is because the dark energy provides the anti-gravitational force in an expanding universe.This makes the solutions in the expanding universe match those in the static universe. We see that the amplitude in (eq. 123)is variable, but at epochs
when the temperature anisotropy of radiation is dominant, which is the
late recombination epochs, the term $\lambda$ can be negleted. So we write
(eq. 123) as:
  \begin{equation}
  \lbrace\Theta -1\rbrace =\frac{3A e^{\phi_{do}}}{\delta\phi_{d}}\cos\alpha
  \end{equation}
  Using the result of (eq. 61), we write:
  \begin{equation}
  \theta\approx\frac{\delta\phi_{d}}{3}
  \end{equation}

The equation for fractional amplitudes  of perturbations in the cold
dark matter that originate at scales smaller than the mean free path of
the photons is:
  \begin{equation}
  \delta_{d} = -\frac{4Ac_{s}^{2}a}{9\varepsilon_{do}}\sqrt{\lbrace 1 + (\frac{\lambda}{\alpha})^{2}\rbrace}e^{-\lambda t}\cos\lbrace\alpha -\arctan\frac{\lambda}{\alpha}\rbrace
  \end{equation}
  and for late recombination epochs, neglecting $\lambda$,  we can write:
  \begin{equation}
  \delta_{d} = -\frac{4Ac_{s}^{2}a}{9\varepsilon_{do}}\cos\alpha 
    \end{equation}
 where \begin{equation} \alpha =\sqrt{\lbrace 4\pi G\varepsilon_{b\gamma o}(1-\frac{1}{a})-k^{2}\varepsilon_{b\gamma o}(\frac{c_{s}^{2}-v_{b\gamma o}u_{o}}{a^{2}})-\frac{\dot{u_{o}}}{a}\rbrace -(\frac{-u_{o}}{2a})^{2}}
  \end{equation}

The above equation shows that the fractional amplitudes of perturbations
in cold dark matter at small scales after equality keep growing with
scale factor, but with a decreasing frequency of oscillations. 
\section{The core-cusp problem:New Theory for Rotation curves of Giant LSB Galaxies}
\label{sec:4}
We know that the discrepancy in the determination of central mass distributions using rotation curves i.e. through astronomical observations and the one from $\Lambda$CDM simulations is known as the core-cusp controversy.This is one of the major crisis in small scale $\Lambda$CDM cosmology.The value of inner mass density power-law slope i.e. $\alpha =0 $ gives the core model,which has approximately constant matter density.This implies that the rotation curve rises in a linear fashion in the core model.In the same way,the value of $\alpha =-1$ gives the cusp model,in which the dark matter density rises in a "spiky"fashion towards the centre.              
We know that the CDM predicts halos that have a high density core.Their inner profile is too steep to be compared to observations. The rotation curves data prefer halo models over those with a cusp \cite{KuziodeNaray2007}. It has been shown that this is true not only for LSB galaxies, but for all galaxies \cite{salucci}. While the quality and quantity of these simulations has improved by orders of magnitude over the years,there is still no "cosmological theory"that explains the distribution of dark matter in galaxies from first principles.A need for an entirely different DM model which naturally produces the cored halos has been proved in \cite{Kaufmann2011}. \\
The amount of luminous matter(stars and galaxies)in many spirals and irregular galaxies is not sufficient to explain the amplitude and shape of rotation curves.This discrepancy is usually interpreted as evidence for the presence of an extended DM halo surrounding the visible region of galaxies \cite{Cassertano1991}\cite{Persic1996}.The inner region of galaxies is $ \lesssim 10$ kpc. As compared to the size of galaxies,the extent of the DM halos estimated using satellite dynamics is $\sim .2-.5 h^{-1}$Mpc \cite{Zaritsky1994}
\cite{Zaritsky1997}.The inner halo region is $r< 30h^{-1}$kpc.\\
\indent The massive dark-matter dominated galaxies,also called the GLSB(Giant low Surface Brightness) galaxies  act as good probes to explain the core-cusp issue.The type of LSB galaxies most commonly studied with regards to the core-cusp controversy are the late type ,gas-rich,dark-matter dominated disk galaxies.The LSB galaxies are attractive candiddates for $\Lambda$CDM predictions in this regard \cite{Blok1997}.This is because the blow-out process ,forwarded as one of the plausible explanations for the presence of large amounts of dark matter in the outer portion of galaxies can't be effective here \cite{Navarro1996}.Also,the observational evidence indicates that the evolution of such galaxies has been very quiescent,and there is little evidence for major merging episodes,interactions or other processes that might have stirred the baryonic and dark matter \cite{Bothun1997} \cite{Impey1997}.Though,earlier it was believed that the disks  of LSB galaxies were everywhere DM dominated,but recently \cite{Lelli2010}have shown that baryons may dominate the dynamics of inner regions of LSBs.Also,that baryons can alter the halos of LSBs \cite{kdenaray2011}. \\
\indent We here propose a new simple explanation based on our model (i.e.the eq. 36 and eq.110) for the rotation curves of certain massive,dark-matter dominated LSB galaxies e.g. NGC 7589.Such galaxies have very large orbital periods which shows that their outer disks have remain undisturbed from very long epochs of time.We know that the baryons collapse in the gravitational potential of the cold dark matter.At epochs of recombination,(see. eq.34-36 sec.~\ref{sec:2})the acceleration of the cold dark matter reverses its sign when the baryon-radiation plasma is about to decouple from the CDM.In this model,the strongly coupled baryon-radiation plasma ejects out of a concentrated region of cold dark matter.The size of the dark matter region is $\lesssim 50$kpc and the outer skeletal region is of the order of $r\sim 10$ kpc.Thus, when the baryon-radiation plasma is just about to flow out of the outermost skeletal regions of cold dark matter,the acceleration of CDM particles undergoe sign reversal and they start accelerating towards the plasma.This happens in the same way for succesive layers of CDM which then start encircling the baryonic matter.We know  from (eq.36) that the acceleration of the dark matter is equal to the gradient of the gravitational potential of CDM.We can see that in the outermost skeletal layers of CDM ,which are the first layers to start moving towards the plasma the baryonic matter will have the lowest acceleration.This is because as per our assumptions in the model assumed here, the outermost skeletal layers of CDM are always sparsely populated and their density follows an exponentially decreasing profile.Therefore ,they have less gradient of gravitational potential as compared to the inner layers of the skeletal region of dark matter.The innermost layers,which are the last ones to decouple from the ejecting baryon-radiation plasma are densely and uniformly populated with cold dark matter particles.So,they have a very high gradient of gravitational potential in these layers.This implies that the acceleration of the CDM particles is greatest in these regions. Also,even these DM particles start to move towards the baryon-radiation plasma ejecting out of them.This is because they reverse their signs at epochs when the plasma decouples from the CDM.Thus we see that the innermost layers of the skeletal structure of CDM which are the last ones to move towards the baryon-radiation plasma have the highest velocities.In the same way, the outermost layer of the skeletal structure of CDM which are the earliest ones to move towards the plasma have the lowest velocities. Also,we know that the plasma,which is the first one to decouple from the dark mater region at early recombination epochs have highest baryon densities.This plasma which are the earliest one to decouple have lowest velocities.This is because they decouple from the outermost skeletal dark matter where the gradient of gravitation potential is weakest. The successive plasma which decouple from the dark matter regions at late recombination epochs will have lower baryon densities.This plasma which successively decouple from inner DM regions have higher velocities.This is because of high gradient of potential in these inner DM regions.The low velocity baryons in the plasma which are the earliest ones to decouple from skeletal DM regions get scattered by the high density faster moving baryons,which leave the inner DM regions at later epochs.These high density baryons decoupled at late epochs of recombination will act as a source for the following baryons of even higher densities,which decouple from the innermost DM regions.These following baryons coming successively from more inner regions of DM have linearly increasing velocities.The reason for its linear increase is the fact that during the epochs of decouplng ,the increase in the baryon densities in the baryon-radiation plasma from early recombination epochs to last recombination epochs is directly related to a proportional increase in the velocity of sound of the plasma i.e. $c_{s}$ from $0$ to $\frac{1}{3}$.Finally,the slowly moving dark matter will start encircling these central sources of baryonic matter.Thus during galaxy formation the baryon desity approximately rises linearly from the nucleus outwards towards the disk.The size of the disk consisting of the baryonic material and dust will extend upto $\sim 10-20$ kpc for a DM region of size of $\sim 50$kpc.This we try to estimate in the paragraph below:\\
We know that the ratio of the density of baryons to that of dark matter at any epoch is:
\begin{equation}
\frac{\Omega_{b}}{\Omega_{d}}=\frac{0.04}{0.23}=0.1739
\end{equation}
We know from (eq.25)that the ratio of baryon to photon density between equality to recombination is:
\begin{equation}
<\frac{\varepsilon_{b}}{\varepsilon_{d}}>=\frac{4}{9<c_{s}^{2}>}-\frac{4}{3}
\end{equation}
where $<c_{s}^{2}>$is the average of the square of the speed of sound. Now,to calculate the average value of the square of the velocity of sound i.e. $c_{s}^{2}$,we write:
 \begin{equation}
 <c_{s}^{2}>=\frac{1}{2}\lbrace\frac{1}{N}+(\frac{1}{N}+\frac{1}{N-1})+(\frac{1}{N-1}+\frac{1}{N-2})+\dots +(\frac{1}{4}+\frac{1}{3})\rbrace
 \end{equation}

 where $\frac{1}{3}$ is the maximum value of $c_{s}^{2}$ i.e. the square of the velocity of sound at very late recombination epochs.These are epochs when baryons have almost fully decoupled from the radiation i.e. $z\sim 1000$.In the above equation,we take the successive means of $c_{s}^{2}$,as it increases from $\frac{1}{N}$ to $\frac{1}{3}$.To start with the values of N are very large i.e. $c_{s}^{2}$ takes very small values in the beginning i.e. at very early recombintion epochs .These are epochs when the baryons just start decoupling from the radiation i.e. $z\sim $ 1100. 
 \begin{equation}
 <c_{s}^{2}>=\frac{2}{2}\lbrace
 \frac{1}{6}+\frac{1}{4}+\dots +\frac{1}{N-2}+\frac{1}{N-1}+\frac{1}{N}\rbrace
 \end{equation}
 We know,
 \begin{equation}
 e=1+1+\frac{1}{2!}+\frac{1}{3!}+\dots 
 \end{equation}
 \begin{equation}
 <c_{s}^{2}>=\lbrace\frac{1}{6}+\frac{1}{5}+\frac{1}{4}+\frac{1}{6}+\dots\rbrace=\lbrace\frac{11}{30}+(\frac{1}{4}+\frac{1}{6}+\dots)\rbrace\sim\lbrace .3666+(e-2)\rbrace\sim 1.0848
 \end{equation}
Thus,we get
\begin{equation}
\vert <\frac{\varepsilon_{b}}{\varepsilon_{\gamma}}>\vert =0.9236
\end{equation}
Thus,we use (eq.134) and  estimate the total extension of the disk for a DM region of 50 kpc in our case as:
\begin{equation}
0.1739\times 50= 8.695
\end{equation}
The space of 8.695 kpc is for the baryons that leave a DM region of $\sim 50$ kpc.To this we add the space occupied by the photons in the strongly coupled baryon-radiation plasma.We know that the radiation in the coupled baryon-photon fluid  decouples from the baryons when 
they start to leave the DM regions at recombination epochs.The photons then travel freely without much scattering upto the present epochs.The space of the photons in the strongly coupled baryon-radiation plasma is:
\begin{equation}
0.9236\times 8.695 =8.03 kpc
\end{equation}
This 8.03 kpc which is the space of the photons in the plasma,and which after decoupling will remain intact with the baryonic space due to inertia even after matter domination.The space left vacant by the photons could be occupied by dusk or high enery cosmic particles.Therefore,the total size of the disk is
\begin{equation}
8.695+8.0307=16.7\approx 17 kpc
\end{equation}
In our case,we take the disk space of the galaxy inside the DM halo consisting of the baryonic matter,dust and high energy cosmic particles to be 20 kpc.At the outer regions of the galactical disk where the dark matter halo starts,the density will first rise exponentially in the innermost halos and then will remain constant.That is the rotation velocity will first follow the profile of (eq.160) in the DM halo region and then rise proportional to the square root of the distance from the galactic centre as in(eq.163).Therefore,according to the mechanism discussed above when the dark matter halos encircle the luminous matter of the galaxies,we get the following structure.The innermost region of the DM halos start from 20 kpc and extends upto 25 kpc.This innermost halo region of $\sim 10$kpc was the outermost skeletal region of the DM expanse before the DM particles reversed their velocity at recombination epochs and start to move towards the baryons very slowly.So,more farther a region before decoupling in the DM expanse,the more inner region it becomes after halo formation.Therefore,the density grows exponentially in this region from the innermost halos upto 10 kpc i.e between  20 kpc and 25 kpc. \\
\indent  The baryons which successively leave the innermost regions of DM will have their velocities icreasing approximately proportional to the increasing velocity of the square of the velocity of sound $c_{s}^{2}$.The baryons which leave from the outermost regions of DM at early recombination epochs,i.e. the ones with the lowest velocities get scattered away by the high velocity baryons coming from the inner DM regions at later epochs.So,the baryons from the successive inner DM regions are scattered lesser and lesser.Therefore,the probability of the baryons remaining in the inner regions of galaxies is proportional to their velocities.In our case,we assume that the clustering of baryons in the inner regions of disk starts from those baryons which leave the DM regions at approximately halfway between early and late recombination epochs i.e.z$\sim 1050$. Thus,inside the galactic disk the rate of increase of the baryonic densities from the galactic distance will be approximately proportional to the rate of increase of baryonic velocities with the distance from the galactic centre.  \\
 The baryons which are the earliest ones to leave the skeletal DM regions have very low velocities .Therefore,they are scattered by the baryons with higher and higher velocities, which successively leave from the inner regions of DM having high gravitational potential.So,we assume that the clustering of baryons will start at epochs which are approximately halfway between epochs of early and late recombination i.e. $z\sim 1050$.To calculate the baryonic densities at $z\sim 1050$,we use
 \begin{equation}
 \varepsilon_{z}=\Omega_{b}\varepsilon_{cr}(1+z)^{3}
 \end{equation} 
 where $\Omega_{b}= 0.04$ and $\varepsilon_{cr} =1.88h^{2}\times 10^{-29} gcm^{-3}$is the critical density of matter today.Here $h=.72$.Thus,we get
 \begin{equation}
 \varepsilon_{b} (z\sim 1050)=0.045\times 10^{-20}gcm^{-3}
 \end{equation}  
 Since,the scaling factor a corresponding to $z\sim 1050 $ is $10^{4}$,the effective densities today will be
 \begin{equation}
 \varepsilon_{eff}=10^{-4}
 \varepsilon_{z}
 \end{equation}
 So,we write
 \begin{equation}
 \nabla^{2}\phi
= 4\pi G\varepsilon_{eff}
\end{equation}
Since,baryons dominate the inner regions of such massive dark matter dominated galaxies upto $r\sim 20$ kpc,where r is the distance from the galactic centre.Therefore for $r\leq 20 $ kpc,we use (eq.142) and (eq.143) to write
\begin{equation}
\frac{v^{2}}{r}=\nabla\phi =4\pi G\varepsilon_{b}(z\sim 1050)\times 10^{-4}
\end{equation}
 Therefore we get
 \begin{equation}
 v=5.9891r
 \end{equation}
 The region between $20\leq r\leq 25$ kpc,which was earlier the outer DM skeletal region of dark matter will now form the innermost region of the DM halo starting from the galactic centre.This is the region from where the baryons decouple at early recombination epochs i.e. $z\sim 1100$,we write for DM densities 
 \begin{equation}
 \varepsilon_{d}(z\sim 1000)
 =\Omega_{d}\varepsilon_{cr}(1+z)^{3}
 \end{equation}
 where $\Omega_{d}=.23$.
We assume the densities to be  increasing exponentially inwards in the skeletal DM regions before decoupling.Thus after the baryons start to decouple and the DM particles reverse their velocities .So,when the DM encircles the inner baryonic region the density increases exponentially in the inner halo region i.e. $20<r\leq 25$ kpc.Its profile is
\begin{equation}
\varepsilon_{d}=\varepsilon_{0}
\exp[\frac{r-20}{5}]
\end{equation} 
where $\varepsilon_{0}=0.29915\times
10^{-20}$gcm$^{-3}$ is the dark matter density at early recombination epochs i.e.$z=1100$.
We adjust for the scaling factor $a\sim 10^{4}$ for recombination epochs and write
\begin{equation}
\varepsilon_{eff}=10^{-4}\varepsilon_{0}
\exp[\frac{r-20}{5}]
\end{equation}
We use (eq.145) and write
\begin{equation}
v=34.53\exp[\frac{r-20}{10}].r^{1/2}
\end{equation} 
The halo region at $r\geq 25$ kpc are the innermost regions of dark matter.These are the DM regions from which the baryons decouple at very late recombination epochs.We know at very late recombination epochs dark matter no longer remains the dominant component.This is due to the fact that the dark energy starts to increase.Therefore,we write
\begin{equation}
\nabla^{2}\phi \approx 0
\end{equation}
Thus for $r\geq 25$ kpc,we write
\begin{equation}
\frac{v^{2}}{r}=\nabla\phi = C
\end{equation}
\begin{equation}
v=C^{1/2}r^{1/2}
\end{equation}
At $r=25$ kpc,$v=284.52$km/s.Therefore,$C^{1/2}=56.905$ km$s^{-1}$kp$c^{-1}$.
Thus,we get for $r\geq 25$kpc
\begin{equation}
v=56.905\times r^{1/2}
\end{equation} 
\begin{figure}
\centerline{\resizebox{10.0cm}{!}{\includegraphics{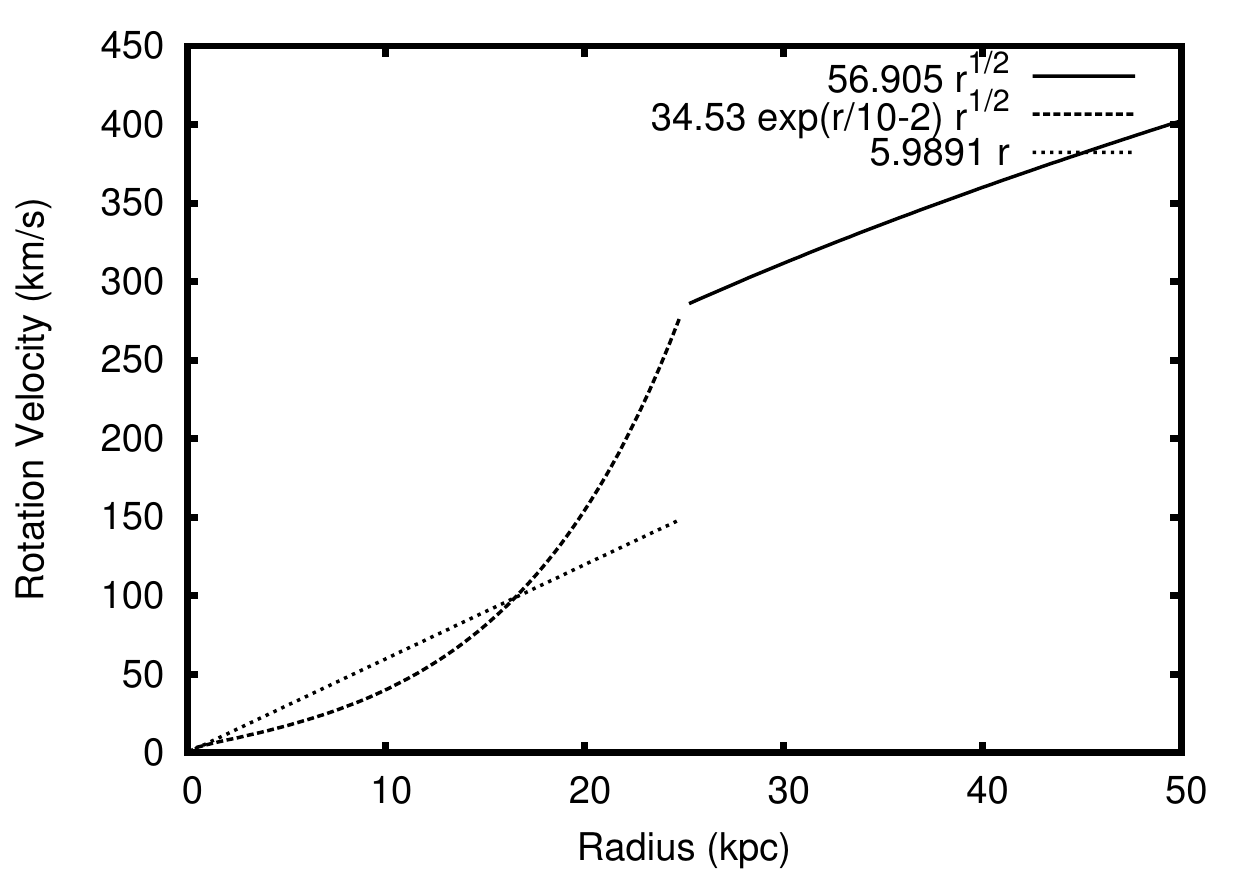}}}
\caption{ The plot shows calculated rotation velocities as a function of distance from the galactic center.}
\end{figure}
\section{Conclusion}
\label{conclusion}
The dynamics of feebly,self-interacting,collisionless,cold dark matter using variable mass formalism of Merscerskii provides a new simple explanation for the observed rotation curves of massive DM dominated LSB galaxies.It also correctly explains the CMB,one of the important pillars of modern cosmology.The L.H.S. of (eq. 2) proves that the cold dark matter can feel some pressure due to its heavy mass and small fluctuations in its velocity due to self-interactions.Thus slowing out of such a feebly self-interacting DM can justify some recent galaxy mergers ,wherein the galaxies outrun dark matter. 
We conclude that this model, produces new results,which relate the oscillations in the CMB anisotropy to the ejection velocity of the baryon -radiation plasma (eq. 31).We know that the height of the second peak in the CMB is particularly sensitive to the baryon-to-photon ratio.Thus,the presence of the ejection velocity which is sensitive to the baryon -to-photon ratio gives a good tool for numerical work ,with regard to exploring the oscillations of CMB in relation to their dependence on the ejection velocity.The assumptions in the model (eq. 27) which ensure the equilibrium of coupled baryon -radiation plasma at all epochs between equality and recombination,  even in
the presence of dark matter is a strong statement for the collisionless nature of CDM .The L.H.S. of (eq. 2) is possible only because the cold dark matter can feel some pressure due to its heavy mass and small fluctuations in its velocity due to self-interactions.The results of (eq. 62) and (eq. 130) represent the Sachs-Wolfe
effect.The (eq. 89) correctly predicts the growth of fractional amplitudes
of perturbations in cold dark matter, which originate between equality
and recombination epochs.For late recombination epochs,it reduces to the standard result $ \delta_{d}\propto a$.The (eq. 59) and (eq. 127) describe the affect
of the perturbations in the cold dark matter on the CMB temperature
anisotropy fluctuations. They also highlight that the dominant component
of CMB temperature anisotropy fluctuations  is at red-shifts of very late
recombination epochs $z\approx 1000$. This is because the perturabtions
in the cold dark matter are minimal at these epochs but not zero as the average of the potential between epochs of equality and recombination has been considered.The
assumptions of (eq. 32) and (eq. 111), assume the growth of a strange form
of energy(dark energy), with negative pressure, which cancels the force
due to the pressure of baryon-radiation plasma. 

These assumptions also
represent the fact that the rate of increase of ejection velocity of
the Baryon -radiation plasma is only due to the gravitational field
of the Cold dark matter.  It can be inferred that the force due to
its pressure can vanish only to some extent due to low pressure in
high baryonic density regions. This is due to early primordial helium
nucleosynthesis in such regions. Since  the baryonic densities can never
be very high before recombination. This is because it will hinder the
hydrogen nucleosynthesis which starts after recombination. Therefore, it
has to be assumed that the primary reason for the vanishing of pressure
of baryon-radiation plasma, at near recombination epochs, is that the
dark energy term with negative pressure starts to dominate at these
epochs.Thus the correct explanation of the Sachs-Wolfe effect and the pridominant CMB temperature anisotropy at late recombination epochs based on this model,strongly argue in favor of the collisionless,feebly self-interacting nature of cold dark matter.
This model gives new results ,which can be important for further exploring the second peak of CMB. More accurate results can be obtained by studying the dynamics
of baryon -radiation plasma using energy momentum tensor of an imperfect
fluid in such a scenario. This will be attempted in a future work. 

\begin{acknowledgements} 
We wish to thank Md. Mahfoozul Haque for valuable advice in formatting of
the manuscript in Latex. 
\end{acknowledgements}

   \end{document}